\documentstyle[12pt]{article}
\baselineskip
\advance\textheight by \topskip
\begin{document}
\begin{flushright}
{SU-ITP-95-2}\\
hep-th/9502069\\
February 13, 1995\\
\end{flushright}
\vskip 0.2cm
 \begin{center}
 {\Large\bf GRAVITY AND GLOBAL SYMMETRIES}\\
 \vskip 1.5 cm
{ \bf  Renata Kallosh$^1$,  Andrei Linde$^1$,  Dmitri Linde$^2$,  and  Leonard
Susskind$^1$}
\vskip  0.2cm
{}$^1$ Department of Physics, Stanford University, Stanford, CA 94305-4060,
 USA\\
\vskip 0.1cm
{}$^2$ California Institute of Technology,
Pasadena, CA 91125, USA

\end{center}
\vskip .6 cm
\centerline{\bf ABSTRACT}
\vskip -.2cm
\begin{quotation}
\baselineskip 16pt
{}~~There exists a widely spread notion  that gravitational effects can
strongly violate global symmetries. If this  is correct, it may lead to many
important consequences. We will argue, in particular,  that nonperturbative
gravitational effects in the axion theory lead to a strong violation of  CP
invariance unless they are suppressed by
an extremely small factor  $g {\
\lower-1.2pt\vbox{\hbox{\rlap{$<$}\lower5pt\vbox{\hbox{$\sim$}}}}\ }10^{-82}$.

{}~~One could hope that this problem disappears if one represents the global
symmetry of a pseudoscalar axion field  as a  gauge symmetry of the
Ogievetsky-Polubarinov-Kalb-Ramond
antisymmetric tensor field. We will show, however, that this gauge symmetry
does not protect  the axion mass from  quantum corrections. The amplitude of
gravitational effects violating global symmetries could be strongly suppressed
by   $e^{-S}$,  where $S$ is the action of a wormhole which may eat the global
charge.  Unfortunately, in a wide variety of theories based on the Einstein
theory of gravity the action appears to be fairly small, $S = O(10)$.

{}~~However, we have found that the existence of wormholes and the value of
their action are extremely sensitive to the structure of space on the nearly
Planckian scale. We consider several examples  (Kaluza-Klein theory, conformal
anomaly, $R^2$ terms) which show that modifications of  the Einstein theory on
the length scale $l {\
\lower-1.2pt\vbox{\hbox{\rlap{$<$}\lower5pt\vbox{\hbox{$\sim$}}}}\ }10
M_P^{-1}$ may strongly suppress violation of global symmetries. We
have found also that in string theory there exists an additional suppression of
topology change by the factor $e^{- {8 \pi^2 \over   g^2}}$.
This effect is strong enough to save the axion theory  for the natural values
of the stringy gauge coupling constant.\\
\end{quotation}

\newpage
\baselineskip 18 pt

\section{\label{1s} Introduction}

The most elegant way to solve the strong CP
violation problem is given by the Peccei-Quinn mechanism \cite{pq}.
This mechanism is based on the assumption that there exists a complex
scalar field
$\Phi(x) \equiv {f(x)\over \sqrt 2} e^{i\theta(x)}$, ~which after spontaneous
symmetry breaking can be represented as
${\phi(x) + f_0 \over \sqrt 2}\, \exp\Bigl({i a(x)\over f_0}\Bigr)$. The
Goldstone field
$a(x)$ (axion) has  the coupling ${a\over 32\pi^2 f_0} F_{\mu\nu}\tilde
F^{\mu\nu}$,\,
similar to the famous $\theta$-term ${\bar\theta\over 32\pi^2} F_{\mu\nu}\tilde
F^{\mu\nu}$.
Nonperturbative effects in QCD lead to appearance of the condensate
$\langle F_{\mu\nu}\tilde F_{\mu\nu}\rangle$,
%%%
and to the effective potential of the axion field  proportional to
$\Lambda^4_{QCD}(1-\cos(\bar\theta + {a\over f_0}))$. This potential has a
minimum at ${a\over f_0} = -
\bar\theta$. In this minimum the terms ${a\over 32\pi^2 f_0} F_{\mu\nu}\tilde
F^{\mu\nu}$ and  ${\bar\theta\over 32\pi^2} F_{\mu\nu}\tilde
F^{\mu\nu}$ cancel each other,  and strong CP violation disappears. This effect
gives the axion a
small mass  $m_a \sim {\Lambda^2_{QCD}\over f_0}$.

In addition to providing a possible solution to the strong CP violation
problem,   invisible  axion field \cite{invisible} is one of  the best dark
matter candidates \cite{ax}. It naturally  appears in
all phenomenological models based on superstring theory \cite{GSW}. Axion
field
 possesses many interesting properties near black holes \cite{Hair}. Finally,
axions may be responsible for a possible existence of wormholes in the
baby-universe
 theory \cite{GS}. Therefore there exists an extensive literature on  axions.
This literature includes at least two different  formulations of the axion
theory,   which are  not completely  equivalent, and several modifications of
these formulations (for a review see \cite{Peccei}).

The axion theory has many problems. First of all, it is not easy to make this
theory  compatible with cosmology. If the spontaneous symmetry breaking
towards a state with $f_0 \not = 0$ occurs after the end of inflation, then the
standard axion model is compatible with cosmological and astrophysical
constraints only if $10^{10}$ GeV
${\ \lower-1.2pt\vbox{\hbox{\rlap{$<$}\lower5pt\vbox{\hbox{$\sim$}}}}\ }f_0{\
\lower-1.2pt\vbox{\hbox{\rlap{$<$}\lower5pt\vbox{\hbox{$\sim$}}}}\ }10^{12}$
GeV \cite{axion}. Recent investigation with an account of cosmological effects
of the axion strings suggests that the upper bound  may be even more tight, so
that the ``axion window'' becomes almost closed, $10^{10}$ GeV
${\ \lower-1.2pt\vbox{\hbox{\rlap{$<$}\lower5pt\vbox{\hbox{$\sim$}}}}\ }f_0{\
\lower-1.2pt\vbox{\hbox{\rlap{$<$}\lower5pt\vbox{\hbox{$\sim$}}}}\ }10^{11}$
GeV \cite{Davis}. On the other hand, if
the spontaneous symmetry breaking occurs during inflation, then the
constraint $f_0{\
\lower-1.2pt\vbox{\hbox{\rlap{$<$}\lower5pt\vbox{\hbox{$\sim$}}}}\ }10^{12}$
disappears \cite{[11]}, but typically it implies
that the Hubble constant at the end of inflation should be sufficiently small,
$H {\ \lower-1.2pt\vbox{\hbox{\rlap{$<$}\lower5pt\vbox{\hbox{$\sim$}}}}\ }10^9$
GeV  \cite{TurWil,LinAxion}.  Inflationary models of this type can be
easily suggested \cite{LinAxion,Hybrid}, but one should keep in mind that not
every
inflationary model satisfies this condition.

Three years ago  it was pointed out  \cite{pqgrav} that  the  axion theory
faces  another difficult problem, which we are going to discuss in this paper.

The standard potential in the axion theory (ignoring small QCD corrections)
is given  by
\begin{equation}\label{1}
V_0(\Phi)=\lambda(|\Phi|^2-f_0^2/2)^2 \ .
\end{equation}
In this approximation the axion is massless due to the global symmetry
$\Phi \to
\Phi \, e^{i\theta}$. However, there are some reasons to expect that
nonperturbative quantum gravity effects do not respect global symmetries.
The simplest way to  understand it is to remember that global charges can be
eaten by black holes, which subsequently may evaporate. One may expect that a
similar  effect can occur because of a nonperturbative formation and
evaporation of ``virtual black holes'' in the presence of a global charge. A
somewhat more developed (even though still very
speculative) approach is based on investigation of wormholes, which may
take a global charge from our universe to some other one. Indeed, it was
claimed in
\cite{GS,AW,CL}  that such effects do actually take place, and  can be
described
by   additional terms (vertex operators) in the effective Lagrangian  which
break the global
symmetry.

 As an example, one may consider the
terms of the following type
\cite{pqgrav}:
\begin{equation}\label{2}
V_g(\Phi) = g_n\, {  |\Phi|^{2m}\, \Phi^{4-2m+n}\over  M_{\rm P}^n} + h.c.
\ ,
\end{equation}
where $g_n$ is some dimensionless constant.  Naively, one could expect that
these
operators should be at least of the fifth  order in
$\Phi$, so that they should be suppressed by $M_{\rm P}^{n}$ in the
denominator, with $n> 0$. The authors of \cite{pqgrav} concentrated
on the simplest (and the most dangerous) term
$g_5  {|\Phi|^{4} (\Phi+\Phi^*)\over M_{\rm P}}$. They have shown that,  for
$f_0
\sim 10^{12}$ GeV, this term  destroys the standard solution of the strong CP
problem. Indeed, this term changes the shape of the effective potential and
moves its minimum away from ${a\over f_0}  = -
\bar\theta$. If  ${a\over f_0}$ changes by more than $10^{-9}$,  the
corresponding
effects of CP violation become too strong. In order to avoid such effects, one
should have an extremely small coupling constant $g$ of the symmetry
breaking operator  $g_5  {|\Phi|^{4} (\Phi+\Phi^*)\over M_{\rm P}}$: \,
$g_5 <10^{-54}$ for $f_0 \sim 10^{12}$ GeV \cite{pqgrav}. Thus, instead of the
problem of explaining why
the angle
$\bar\theta$ in the theory of strong interactions is smaller than $10^{-9}$, we
must explain now why some other parameter is smaller than $10^{-54}$. This
does not look like a fair trade!

In fact, the situation is even more complicated. The idea to consider only the
terms containing $M_{\rm P}$ in the denominator  was based on the  assumption
that the quantum gravity effects should be suppressed in the limit $M_{\rm P}
\to
\infty$. Indeed, the
$n$-loop quantum gravity corrections contain factors   $M_{\rm P}^{-n}$.
However, the effects
we are interested in are {\it nonperturbative}.  These effects may give rise to
vertex operators of the type of $g_1 M_{\rm P}^3 (\Phi+\Phi^*)$, or other
operators
which do not contain $M_{\rm P}$ in the denominator
\cite{AW,CL}.   The way to see it is, e.g.,  to consider the averages of the
type $M_{\rm P}^3\langle  (\Phi+\Phi^*) \rangle$. One can show that in the
presence of wormholes such terms do not vanish, but they are suppressed by the
same exponential factor $e^{-S}$ as the terms $\langle {|\Phi|^{4}
(\Phi+\Phi^*)\over M_{\rm P}}\rangle$ \cite{AW}. Here $S$ is the action of a
wormhole which can ``eat'' a unit of a global charge associated with the field
$\Phi$. This implies that if  the effective vertex operators $g_5 {|\Phi|^{4}
(\Phi+\Phi^*)\over M_{\rm P}}$   appear in the theory due to nonperturbative
gravitational effects, one may expect   that the operators $g_1 M_{\rm P}^3
(\Phi+\Phi^*)$ should appear as well, with a comparable coupling constant, $g_1
\sim g_5  \sim e^{-S}$.

The vertex operator  $g_1 M_{\rm P}^3 (\Phi+\Phi^*)$
is most dangerous for the axion physics. One can easily show, by analogy with
\cite{pqgrav},  that this term   leads to a strong CP violation unless $g_1 {\
\lower-1.2pt\vbox{\hbox{\rlap{$<$}\lower5pt\vbox{\hbox{$\sim$}}}}\ }10^{-82}
{f_0\over 10^{12} \rm GeV}$. This constraint is almost thirty orders of
magnitude stronger that the constraint following from the investigation of the
operators $g_5  {|\Phi|^{4} (\Phi+\Phi^*)\over M_{\rm P}}$. For comparison of
our constraint with the results of our future calculations of the wormhole
action it is convenient  to express this constraint (for $f_0 \sim 10^{12}$
GeV) in the form $g_1{\
\lower-1.2pt\vbox{\hbox{\rlap{$<$}\lower5pt\vbox{\hbox{$\sim$}}}}\ }e^{-189}$.

Note, that this constraint depends on the value of $f_0$, but not too strongly:
it is proportional to $f_0$. Thus, for $f_0 \sim 10^{10}$ GeV one should
have $g_1{\ \lower-1.2pt\vbox{\hbox{\rlap{$<$}\lower5pt\vbox{\hbox{$\sim$}}}}\
}e^{-193}$, whereas for    $f_0 \sim 10^{19}$ GeV our constraint is $g_1{\
\lower-1.2pt\vbox{\hbox{\rlap{$<$}\lower5pt\vbox{\hbox{$\sim$}}}}\ }e^{-172}$.
In what follows we will suppose, for definiteness, that $f_0 = 10^{12}$ GeV,
even though, as we have already emphasized, $f_0$ in the axion theory may be
either two orders of magnitude smaller, or much greater than $10^{12}$ GeV.

Similar arguments are valid for  other  theories possessing global symmetries.
For example, it was shown in  \cite{Textures} that the theory of cosmic
textures may work only if the constant $g$ in the term $g_5  {|\Phi|^{4}
(\Phi+\Phi^*)\over M_{\rm P}}$ is extremely small: $g_5<10^{-91}$.
One can easily show that this constraint becomes even much stronger if one
takes into account the above-mentioned terms linear in
$\Phi$: $g_1 < 10^{-103} \sim e^{-237}$.

The same situation appears in the so-called ``natural inflation'' model
\cite{Natural}. In this model it is assumed that the effective potential has
the form (\ref{2}) with $f_0 {\
\lower-1.2pt\vbox{\hbox{\rlap{$>$}\lower5pt\vbox{\hbox{$\sim$}}}}\ }M_{\rm P}$,
and then the Goldstone field, just like
the axion field, acquires mass $\Lambda^2/f_0 \sim 10^{13}$ GeV. This can
be achieved in a natural way, e.g.,  for $\Lambda \sim 10^{16}$ GeV and $f_0
\sim M_{\rm P}$. However, the term  $g_1 M_{\rm P}^3 (\Phi+\Phi^*)$ will
destroy this nice picture unless the coupling $g_1$ is extremely small, $g_1{\
\lower-1.2pt\vbox{\hbox{\rlap{$<$}\lower5pt\vbox{\hbox{$\sim$}}}}\ }
10^{-6}$. As compared with the constraints on the models of axions and
textures, this condition looks relatively mild. Still, the existence of this
additional constraint is rather disappointing. (On the other hand, it would be
quite encouraging to find a natural mechanism which would lead to the
gravity-induced terms  $\sim g_1 M_{\rm P}^3 (\Phi+\Phi^*)$ with $g_1{\
\lower-1.2pt\vbox{\hbox{\rlap{$<$}\lower5pt\vbox{\hbox{$\sim$}}}}\ }10^{-6}$,
since some terms breaking the global symmetry are necessary in this
scenario.)

As we already mentioned, the main reason why quantum gravity may break
global symmetries is associated  with the possibility that the global charge
can be eaten by wormholes (or virtual black holes) and taken away from our
universe. It is commonly believed, however, that local charges, such as
electric or magnetic one, cannot disappear, and therefore quantum gravity
does not break local symmetries. The reason can be easily understood if one
think about electric (or magnetic) charges falling into black holes. Because of
the Gauss law, the  flux of electric field cannot disappear when the charge
falls into a black hole. Charged black holes cannot evaporate entirely and take
the electric charge away from our universe. Instead of that,  they eventually
form charged extreme black holes which do not evaporate any further.

It would be very tempting to use a similar mechanism to save the axion theory.
Indeed, it is well known that the theory of a massless pseudoscalar axion field
in a topologically trivial space is equivalent to the theory of an
antisymmetric  tensor gauge field $b_{\mu\nu}$. This field, which  was
introduced by Ogievetsky and Polubarinov  and later also by Kalb and Ramond
\cite{Ogiev}, naturally
appears in string theory.   It is related to
 the field $a$ by the duality transformation. This suggests an idea that if
one  formulates the axion theory in terms of the antisymmetric {\it gauge}
field, then the low mass of the axion will be protected not by global but by
local (i.e. gauge) invariance, and it will not be destroyed by quantum gravity.

Of course, one may immediately argue that this cannot work. Indeed, if the
axion  charge can disappear in the standard formulation of the theory, its
disappearance may have an adequate description in the theory of an
antisymmetric tensor field.
However, this argument has some caveats, since in fact these two theories are
not  completely  equivalent at the  quantum level. For example, conformal
anomaly associated with the pseudoscalar field differs from the conformal
anomaly in the theory of the antisymmetric tensor field \cite{NONEQUIV}.

There is another problem which appears to be much more important in the context
of our discussion.  Whereas a massless antisymmetric tensor field  in a
topologically trivial space can be converted into a pseudoscalar, not all
pseudoscalars can be
replaced by antisymmetric tensors. The necessary condition is that
the Lagrangian  of a pseudoscalar  field should depend only on its
derivatives.

Indeed, if one starts with the theory of the antisymmetric tensor gauge field,
duality transformation   between the  field strength of the antisymmetric
tensor  field $ \epsilon^{\mu\nu\lambda \delta}  \partial_{ \nu } b_{
\lambda \delta  }$ and derivative of the  pseudoscalar  field
$\partial^{\mu}a$
exists and can be used to prove the equivalence of these two
theories  \cite{NT}. Duality transformation is always possible from the
$b_{\mu\nu}$-side
to $a$-side and one ends up
with the theory of the  massless pseudoscalar  field with derivative
coupling only. Vice versa, if one starts with the theory of the  massless
pseudoscalar  field with derivative
coupling only, one can use the duality transformation and have
an equivalent $b_{\mu\nu}$-version of the theory.

If the effective action  depends on  the pseudoscalar  field $a$ without
derivatives  (and this is the case if the axion field has a small mass!), no
clear
information about the relation of this theory to the $b_{\mu\nu}$-theory was
available.
To clarify the relation between $b_{\mu\nu}$-  versus
$a$-theories one should
understand how one can describe appearance of the small mass $m_a  \sim
{\Lambda^2_{QCD}\over f_0}$ in terms of the gauge theory of the field
$b_{\mu\nu}$. Is
this effect possible at all, or is it associated with some kind of gauge
symmetry
breaking?

We have analysed this question and found that the axion mass generation can be
consistently described in terms of the antisymmetric tensor   field, and that
this effect does not
involve any gauge symmetry breaking. This effect is quite interesting in
its own terms, independently of the initial goal of our investigation. It
provides a generalization of the phenomenon studied by Polyakov in
three-dimensional compact QED, where the massive scalar excitation appears in
presence of monopoles \cite{Pol}. In our case the mass of the antisymmetric
tensor field appears because of its interaction with the usual QCD instantons.
The deep physical reason why the antisymmetric tensor   field can acquire mass
without breaking gauge invariance is that both the electromagnetic field in
$d=3$ space-time and the antisymmetric tensor field in $d=4$ space-time have
only one degree of freedom. Thus, the condition of transversality for these
theories, which typically protects  excitations from becoming massive, does not
apply to   {\it physical} degrees of freedom in these theories. We will
describe this effect
 in Section 2 of this paper.

Even though there is no reason to expect that the
Peccei-Quinn symmetry is protected by the gauge invariance of the antisymmetric
tensor field, one may still hope that the symmetry violation should be very
small for some other reason. For
example, it is not so easy for a quantized axion charge to be squeezed into a
wormhole or a black hole: They should be large enough to eat a unit of the
axion
charge. In the language of Euclidean quantum gravity this translates into the
question of what is the Euclidean action of the wormhole which could eat a unit
of the
quantized axion charge. If this action $S$ is large enough, then one could
expect that the violation of the Peccei-Quinn symmetry is strongly suppressed
by a factor $e^{-S}$.  Our estimates of the effects related to the term $g_1
M_{\rm P}^3 (\Phi+\Phi^*)$ indicate that  in order to suppress dangerous
effects of global symmetry violation in the axion theory one should have the
wormhole action $S  {\
\lower-1.2pt\vbox{\hbox{\rlap{$>$}\lower5pt\vbox{\hbox{$\sim$}}}}\ }190$.

Wormholes which could eat the global charge have been first discovered by
Giddings and Strominger  in the formulation of the axion theory in terms of the
antisymmetric
tensor field \cite{GS}.  The simplest of their solutions corresponds to the
fixed value of the radial component of the scalar field.    The pseudoscalar
representation of  the Giddings-Strominger wormhole    was obtained by
Kimyeong Lee \cite{Lee}.  In Section 3 of this paper we will re-derive their
expression for the wormhole action. Our result for the value of the action of
the wormhole configuration without the boundary at the wormhole neck
 coincides with the result obtained by Giddings, Strominger and Lee.
However, we point out that if one takes into account boundary terms including
the contribution of  the boundary at the wormhole throat, the action becomes
about three times smaller.
In any case, the action is proportional to
${M_{\rm P}\over f_0}$, which is as large as $10^7$ for the  axion theory with
$f_0
\sim 10^{12}$ GeV. Thus could suggest that for the axion theory one does not
have any
problem whatsoever since the symmetry violating effects will be suppressed by
the
factor $\sim 10^{-10^7}$, which is more than enough to explain why $g_1 {\
\lower-1.2pt\vbox{\hbox{\rlap{$<$}\lower5pt\vbox{\hbox{$\sim$}}}}\ }10^{-82}$.

Unfortunately, this  attitude proves to be too optimistic. As was first pointed
out in
\cite{AW}, in realistic models of the axion field  the radial component $f(x)$
of the axion field  on the wormhole solutions does not remain  equal to $f_0$.
Near
the wormhole throat this field typically acquires some value of the order of  $
M_{\rm P}
\gg f_0$. A detailed investigation of solutions with an account taken of the
spatial
dependence of  $f(x)$ was performed   by Abbott and Wise \cite{AW} and by
Coleman and  Lee \cite{CL} for the case without spontaneous symmetry breaking.
The corresponding Euclidean action which was found in these papers linearly
diverged on extremely large length scales. It was argued that despite the
action is
infinite, wormholes do lead to charge nonconservation and global symmetry
breaking
since the corresponding effects appear on a relatively small scale, where the
large
scale behavior of the wormhole solutions is irrelevant.

Unfortunately, the most interesting case of the theories with spontaneous
symmetry breaking was only briefly mentioned in  \cite{AW,CL}.  It required
some additional work to obtain results in a form in which one could compare
them with the expectations expressed in \cite{pqgrav}. Perhaps  this was the
reason why the authors of ref. \cite{pqgrav} did not make any attempt to use
the results obtained in
\cite{GS,Lee,AW,CL}.

In Section 4 of this paper we describe wormhole solutions in several different
theories with spontaneous
symmetry breaking. Whereas in some cases we could obtain important information
about these solutions by the methods of ref. \cite{AW}, in general it was
necessary to use numerical calculations. These calculations were extremely
tedious,
especially for the axion theories with $f_0 \sim 10^{12}$ GeV. The results
which we
obtained are in a qualitative agreement with the expectations of \cite{AW,CL}.
We have shown that for a very wide class of potentials the action is
finite and to a good accuracy  is given by a simple expression $S \sim
\ln{M_{\rm P}\over f_0}$. This means that  if the global symmetry breaking is
suppressed by $e^{-{S}}$,
this suppression is approximately given by the factor ${f_0\over M_{\rm P}}$.
This
factor is of the order of $10^{-7}$ for the axion theory with $f_0 \sim
10^{12}$ GeV, and it is of the order of $10^{-3}$ for the texture theory with
$f_t \sim 10^{16}$ GeV. This is clearly insufficient to save the axion and the
texture theory.

In fact, the situation becomes even more complicated if one takes into account
that,
according to \cite{AW,CL},  the symmetry breaking vertex operators
$g_{n}\,
{  |\Phi|^{2m} \Phi^{4 - 2m +n}\over  M_{\rm P}^{n}} + h.c.$ are suppressed
only by
the part of the
Euclidean action $S$ corresponding to integration in a small vicinity of the
throat of
the wormhole. In this case suppression of the global symmetry
breaking in the theory $-m^2 |\Phi|^2 + \lambda|\Phi^2|^2$ practically
disappears.

One could expect that this result should be strongly model-dependent. Indeed,
we have found
that near the wormhole throat the field $f(x)$ typically becomes as big as
$M_{\rm P}$. In
certain theories, such as the theory of superstrings or supergravity, the
effective
potential  may acquire large additional terms, which exponentially grow at
large
$f$. In such theories the behavior of the wormhole solutions near the throat
becomes quite different from the one envisaged in \cite{AW,CL}. Here our use of
numerical methods was absolutely crucial. The results, however, have not been
very
encouraging: Even if one considers the effective potentials growing at large
$f$ as
fast as $\exp{500 f\over M_{\rm P}}$, the resulting Euclidean action   remains
quite small. Thus, one cannot make the violation of the global symmetries small
by changing the effective potential of the scalar field in any reasonable way.

Fortunately, during our investigation we have found several ways to fix this
problem. First of all,  when we make the effective potential more and
more steep (keeping its minimum at $f_0$), the corresponding action tends to
increase towards the very large action of the Giddings-Strominger-Lee wormhole.
We  have found that the main reason why it happens is an increase of the size
of the wormhole: The action  is (approximately) proportional to the square of
the radius of the wormhole throat $R(0)$, $S \sim  M^2_{\rm P} R^2 (0)$.
According to our calculations, increasing of the minimal
radius of the throat  just a few times as compared with $M_{\rm P}^{-1}$ can
make the ``natural
inflation'' scenario viable.  The situation with axions and textures is more
complicated,
but still these two theories can be saved by an increase of the radius of the
throat up to about $10 M_{\rm P}^{-1}$ or $15 M_{\rm P}^{-1}$.

 It is very difficult to increase this radius by making the effective potential
steep. However, there may be some other reasons why the wormhole throat cannot
be small. For example, in string theory the effective ``minimal length'' may be
somewhat greater than the Planck length $M_{\rm P}^{-1}$.   Our investigation
contained in Section 5 indicates that the size of the wormhole throat can be
very large in Kaluza-Klein theories with a sufficiently large radius of
compactification.  This suggests that the axion theory can be quite viable in
the context of a  theory in which the  gravitational effects on the length
scale   $l {\
\lower-1.2pt\vbox{\hbox{\rlap{$<$}\lower5pt\vbox{\hbox{$\sim$}}}}\ }10\, M_{\rm
P}^{-1}$ cannot be described by  the standard Einstein theory of gravity in a
four-dimensional space-time. Another example pointing in the same direction
is related to ${\cal R}^2$ corrections and effects of conformal anomaly. We
show that an account taken of conformal anomaly (even if the corresponding
terms are relatively small) may lead  to disappearance of the wormhole
solutions.  We also discuss the observation made in \cite{GS,CM} which suggests
that there are no wormhole solutions in certain  versions of the string theory.

In Section 6 we discuss a possibility of an additional strong suppression of
the wormhole effects because of the Gauss-Bonnet term  ${\gamma\over 32\pi^2}
{{{}^*\cal R}}  {{{}^*\cal R}}$. This term  does not change any observational
consequences of the  Einstein theory, but it tends to suppress transitions with
the change of topology. Similar terms appear in the heterotic string theory
with the coefficient proportional to $\alpha'$. We show that these terms may
suppress the wormhole effects by the factor $e^{-{8\pi^2\over g^2}} = e^{-\pi{
M^2_{\rm P}\over M^2_{\rm str}}} $, where $g$ is the gauge coupling constant,
$M^2_{\rm str}$ is the stringy mass scale. This result is very similar to the
standard result $e^{-{8\pi^2\over g^2}}$ for the suppression of the instanton
effects in QCD. The possibility to have this   suppression
factor  smaller than $e^{-190}$  is quite consistent with the present picture
of stringy phenomenology.   This suppression becomes even stronger if one adds
the usual part of the action to the topological contribution discussed above.

Our results and conclusions are summarized in Section 7. A considerable part of
our results is based on numerical investigation of differential equations for
wormholes. In many cases it was impossible to solve these equations using
standard numerical recipes \cite{PTVF}. In Appendix we describe an improved
method which we have used in our work.

\

\section{\label{2s} Axion Theory. Pseudoscalar versus antisymmetric tensor.}

\subsection{\label{2.1}Pseudoscalar formulation of the axion theory}

The simplest version of  the pseudoscalar axion theory \cite{pq} adds to the
standard model Lagrangian ${\cal L}_{SM}$ the
following terms
\begin{equation}
{\cal L} =  {1\over 2} (\partial _\mu a )^2  +  \Bigl( \bar
\theta + {a  \over
f_0} \Bigr){g^2
\over 32
\pi^2}  F_{\mu\nu}^a \, \tilde F^{a \mu\nu}  \ .
\end{equation}
For  a review of various models see \cite{Peccei}. Peccei-Quinn global $U(1)$
symmetry
\begin{equation}
a(x) \rightarrow a(x) + C
\end{equation}
is broken spontaneously when the operator $ F_{\mu\nu}^a \, \tilde F^{a \mu\nu}
= {1\over 2} \epsilon_{\mu \nu \lambda \delta} F^{\mu\nu}  F^{\lambda \delta}$
has a non-vanishing vacuum expectation value $\langle F_{\mu\nu}^a \; \tilde
F^{a
\mu\nu}\rangle$. The  term $F_{\mu\nu}^a \, \tilde F^{a
\mu\nu} $ can be represented as a total derivative, $F_{\mu\nu}^a \, \tilde
F^{a
\mu\nu}  = \partial_\mu  K^\mu$, where
\begin{equation}
K_\mu (x) = \epsilon_{\mu\alpha\beta\gamma} A^{\alpha}_a ( F^{ \beta \gamma}_a
-{g\over 3}  f_{abc} A^{ \beta}_b A^{  \gamma}_c) \ .
\label{CS}\end{equation}
The fact that  $F_{\mu\nu}^a \, \tilde F^{a \mu\nu}$ is a total derivative is
not sufficient for providing
the symmetry since the action is not invariant in presence of instantons.  The
variation of the action with an account taken of the surface terms is
\begin{equation}
\delta S_{eff} =   {C  \over f_0}\,  {g^2 \over 32
\pi^2} \int d^4 x F_{\mu\nu}^a \, \tilde F^{a \mu\nu}  = {C  \over
f_0}\, {g^2 \over 32 \pi^2} \int d^4 x \partial_\mu\,  K^\mu \neq 0 \ .
\end{equation}
Spontaneous breaking of the  global $U(1)$ Peccei-Quinn symmetry allows to
generate a
potential for the axion field. This gives the axion   a small mass
$m_a^2\sim  {m_\pi f_\pi\over f_0^2} \sim  {\Lambda_{QCD}^4 \over f_0^2}$. The
lightness of the axion is provided by the fact that $f_0 \gg \Lambda_{QCD}$.

Note, that the Peccei-Quinn symmetry is global. If the axion would not interact
with non-Abelian fields, this symmetry could be promoted to a gauge one.
Indeed, one can introduce an Abelian vector field \cite{Hair} and replace
everywhere  $ \partial _\mu a$  by $ \partial _\mu a+ e
A_\mu$. Then the global PQ symmetry becomes the local one,
\begin{eqnarray}
a &\rightarrow& a + C(x) \ , \nonumber\\
A_\mu &\rightarrow& A_\mu - \partial_\mu C(x) \ .
\end{eqnarray}
However, in the presence of a non-Abelian gauge coupling, which is an essential
part of all realistic axion models,  one cannot promote the global PQ symmetry
to the local one.   Indeed, the non-Abelian coupling can be represented  in the
form  $\partial_\mu a\,  K^\mu$. After the promotion of the global PQ symmetry
to
the local one  we would obtain the
term   $ e A_\mu K^\mu$ in the action.
This term would violate the  non-Abelian gauge symmetry since the variation of
the Chern-Simons term $K_\mu$ vanishes only when it is coupled to the
longitudinal part of $A_\mu$ but not to the transverse part of it.

 Still it is possible to represent the global PQ symmetry as a local one if one
goes to a dual formulation of the axion theory, in which the axion pseudoscalar
field $a$ is represented by the antisymmetric tensor field $b_{\mu\nu}$. This
version of the axion theory naturally appears in the context of string theory
\cite{GSW}.

\subsection{\label{2.2} Dual version of the axion theory with the non-Abelian
coupling in the broken symmetry phase}

The non-interacting two-index antisymmetric tensor gauge field
$b_{\rho\sigma}$
with the gauge symmetric action
${1\over 2} (\partial _{[\nu} b_{\rho\sigma]} )^2$
is equivalent
to a non-interacting massless pseudoscalar field  with the action ${1\over
2}(\partial_\mu a)^2$.
 The   simplest way
to see this is to perform a duality transformation of the classical action.
This leads to a replacement of
$\partial^\mu a$ by a pseudovector  $ {1\over 2} \epsilon^{\mu\nu \rho\sigma}
\partial _{[\nu} b_{\rho\sigma]}$ which is dual to the tree tensor field
strength
$\partial _{[\nu} b_{\rho\sigma]}$ of the 2-form field $b_{\rho\sigma}$.
Another way to see this equivalence is to perform
gauge fixing of the gauge symmetry $\delta b_{\rho\sigma}= \partial_{[\rho}
\Lambda_{\sigma]}$,  $\delta \Lambda _\sigma = \partial_\sigma \Lambda$,  which
requires two generations of ghosts.
In addition to 6 $b_{\mu\nu}$-fields one gets 8 anticommuting ghosts in first
generation and 3 commuting ones in the second generation. This provides a net
number of propagating degrees of freedom equal to one commuting field.

The equivalence of
the massless pseudoscalar and massless antisymmetric tensor field   coupled to
the non-Abelian gauge field  has been shown in \cite{NT} starting with the
first-order type action using duality transformation. In what follows we will
perform this duality transformation for the theory  which describes
 the complex scalar theory with the axion $\theta$ coupled to the  non-Abelian
vector
fields, \begin{equation}
{\cal L} = |\partial_\mu \Phi|^2 - V(|\Phi|)   - \partial_\mu \theta\;
\Omega^\mu  \ .
\end{equation}
Here
\begin{equation}
\Omega_\mu (x) = {g^2
\over 32
\pi^2 }\, \epsilon_{\mu\alpha\beta\gamma}\, A^{\alpha}_a\, ( F^{ \beta
\gamma}_a
-{g\over 3}  f_{abc} A^{ \beta}_b A^{  \gamma}_c) \equiv    {g^2
\over 32
\pi^2  }  K_\mu\ .
\end{equation}

One can also write this Lagrangian as
\begin{equation}\label{MyTheta}
{\cal L}_{\theta} = {1\over 2} f ^2  (\partial_\mu  \theta )^2    -
\partial_\mu \theta
\; \Omega^\mu
+  {\cal L}(f) \ ,
\end{equation}
where
\begin{equation}
{\cal L}(f)
 \equiv {1\over 2} (\partial_\mu  f )^2 - V(f) \ ,
\end{equation}
and
 \begin{equation}
\Phi = {\phi(x)
+ f_0 \over \sqrt 2}\, \exp\Bigl({i a(x)\over f_0}\Bigr) \equiv  {f(x)
 \over \sqrt 2}\, \exp\Bigl({i \theta (x)}\Bigr) \ .
\end{equation}

However, it is useful to start with a more general Lagrangian which depends
both on the pseudoscalar $\theta$ and on some pseudovector field $H_\mu$,
\begin{equation}\label{Htheta}
{\cal L}_{\theta, H} = i \partial_\mu \theta  (H^\mu + i  \Omega^\mu) + {1\over
2}
  f^{-2}   H_\mu  H^\mu +  {\cal L}(f) \ .
\end{equation}
One can solve the field equations for $H_\mu$,
\begin{equation}
H_\mu = -i  f^2 \partial_\mu \theta \,      \,  \ ,
\end{equation}
and  on shell for $H_\mu$ the Lagrangian acquires the form
 (\ref{MyTheta}),  which describes the pseudoscalar axion
 field $\theta$.

On the other hand,  one can vary   ${\cal L}_{\theta , H}$ (\ref{Htheta}) over
$\theta (x)$ and obtain
the constraint on $H_{\mu}$,
\begin{equation}
\partial _\mu H^\mu + i  {g^2
\over 32
\pi^2  }   F_{\mu\nu}^a \, \tilde F^{a \mu\nu} = 0 \ .
\end{equation}
The solution to this constraint is
\begin{eqnarray}
H^{\mu} &=& -{i\over 2} \epsilon^{\mu\nu\rho\sigma} H_{\nu\rho\sigma} \ ,
\nonumber\\
\nonumber\\
H_{\nu\rho\sigma} &=& \partial _{[\nu} b_{\rho\sigma]} +   {g^2
\over 32
\pi^2  }  A^a_{[\nu} (F_{a \rho\sigma]} - {g\over 3} f_{abc} A^b_\rho
A^c_{\sigma]}) \ ,
\label{con}\end{eqnarray}
where $\partial _{[\nu} b_{\rho\sigma]}$ is the field strength of the 2-form
field.

The Lagrangian which follows from (\ref{Htheta}) describes the antisymmetric
massless field $b_{\mu\nu}$
interacting with the non-Abelian vector fields as well as with the radial
component of the scalar field,
\begin{equation}
{\cal L}_{b_{\mu\nu}}= {1\over 2}  f  ^{-2}  H_{\mu\nu\lambda}^2+  {\cal L}(f)
\ .
\label{dual}\end{equation}

 {\it There are no global symmetries in the dual
version of the axion theory, but there are two types of local symmetries}, the
Maxwell-type gauge symmetry of the 2-form field,
\begin{equation}
\delta b_{\mu\nu} = \partial _{[\mu} \Lambda_{\nu]} \ ,  \qquad\delta
\Lambda_\nu = \partial _\nu \Lambda \ ,
\label{max}\end{equation}
and the Nicolai-Townsend-type non-Abelian gauge symmetry \cite{NT},
\begin{equation}
\delta A_\mu^a = \nabla_\mu^{ab} \Lambda_b\ , \qquad  \delta b_{\mu\nu}= 2
{g^2
\over 32
\pi^2  }
\Lambda^a \partial_{[\mu} A_{a\nu]} \ .
\label{nt}\end{equation}

Will these local symmetries of the dual version allow us to avoid the problem
which
destroys the nice features of the axion theory in the standard version? To
address this issue we will first study the origin of the light axion mass
generated by QCD instantons in the dual version of the theory.\footnote{We are
grateful to M. Dine for the suggestion to investigate this problem.}

\subsection{ \label{2.3} How does the axion become massive in the dual theory?}

The possibility that a scalar or  pseudoscalar particle can acquire a
non-vanishing mass due to non-perturbative effects  does not look very
surprising.  However, in the  dual formulation of the axion theory  the axion
is massless because of a gauge symmetry
(\ref{max}). It is commonly believed that gauge symmetry protects massless
particles from becoming massive. This would solve all our problems.  However,
if it were true, then the axion field in its dual formulation would not get its
mass $m_a \sim  {\Lambda_{QCD}^2\over f_0}$. Therefore before going any further
we must first resolve this puzzle.

With an account of QCD instantons the effective action of the pseudoscalar
field
acquires a potential,
\begin{equation}
{\cal L}_a = {1\over 2} (\partial _\mu a)^2\, {f^2 \over f^2_0}  +
\Lambda^4_{QCD} \cos {a \over f_0}\ .
\end{equation}
Starting with this action one cannot make a transition to the dual version of
the theory with the $b_{\mu\nu}$\,-field using any of the methods discussed
above. For example,  one cannot perform duality transformation of the action
starting with the action depending both on $\partial_\mu a$ and $H_\mu$ as we
did before. Indeed,
 there is now a term in the pseudoscalar version of the theory  depending on
$a$
rather than on $\partial_\mu a$  and the procedure does not work. If we will
try to go from the side of the gauge invariant  $b_{\mu\nu}$\,-theory and
perform duality transformation, we will get no terms
depending on $a$, but only terms depending on  $\partial_\mu a$.

Does the impossibility to perform duality transformation
mean that the dual version of the axion theory is incapable of explaining the
mechanism of generating a small axion mass? Will the same mechanism protect
axion from getting very heavy?

To understand the effect at the level which is more subtle than just performing
duality transformation over the effective action we will first go back   to the
Polyakov model of compact QED in $d=3$ \cite{Pol}. In this theory the massless
scalar is dual to the massless Abelian vector $A_\mu $ and not to the
$b_{\mu\nu}$\,-field.

Polyakov's  main idea  is that the existence of magnetic monopoles, which are
instantons in $d=3$, changes the behavior of the correlators: due to instantons
there are no massless transverse excitations in the system. Instead there is an
excitation corresponding to a longitudinal component of a gauge invariant
operator, which may  be interpreted as  a massive scalar.  The effect has a
non-perturbative nature and does not violate gauge symmetry.

We will show that something very similar happens with the antisymmetric tensor
field in our $d=4$  theory. Let us first describe the situation in $d=3$
\cite{Pol}. One starts with the dual to the vector field strength,
\begin{equation}
H_\alpha = {1\over 2} \epsilon _{\alpha \beta \gamma} F_{\beta \gamma} \ .
\end{equation}
If one allows certain magnetic charge density in the system of the form
$\rho(\vec x) = \sum q_a \delta(\vec x -\vec x_a)$, in the quasiclassical
approximation this would correspond to
\begin{equation}
H_\mu = {2\pi i k_\mu\over k^2} \rho(k) \ .
\end{equation}
The main steps in the calculation of the correlator of two gauge invariant
operators
$H_\mu$ are the following. The bare part without monopoles is
\begin{equation}
\langle H_\mu (k)  H_\nu (-k)\rangle ^{(0)}= e^2
\Bigl(\delta_{\mu\nu} - {k_\mu k_\nu \over k^2} \Bigr) \ .
\end{equation}
This expression corresponds to the free photon propagator coming from the
action
${1\over e^2} F^2_{\mu\nu}$. In addition, there is a contribution from the
instantons such that the full correlator is
\begin{equation}
\langle H_\mu (k)  H_\nu (-k)\rangle = \langle H_\mu (k)  H_\nu (-k)\rangle
^{(0)} + (2\pi)^2 {k_\mu k_\nu \over k^4}\  \langle\rho(k) \rho(-k)\rangle \ .
\end{equation}
The correlator of charge densities is calculated from the generating functional
for the charge density of the plasma and is given in \cite{Pol},
\begin{equation}
Z(\eta) = \langle e^{i \int d^3 x \eta(\vec x)\rho (\vec x)}\rangle = {1\over
Z(0)} \int {\cal D} \chi\, \exp \left \{ -\left ({e\over 2\pi}\right )^2 \int
\Bigl[(\nabla (\chi - \eta))^2 - M^2 \cos \chi \Bigr]\right \} \ ,
\end{equation}
where $M^2 = \left ({2\pi\over e}\right )^2 \exp^{ -{\rm C \over e^2}} $,  $C=
O(1)$ is some constant.
The second variational derivative over $\eta$ gives the expression for the
correlator of $\rho$,
\begin{equation}
\langle\rho(k) \rho(-k)\rangle = \left ({e\over 2\pi}\right )^2  \left(k^2 -
{k^4\over  M^2 + k^2}\right) = \left ({e\over 2\pi}\right )^2
{M^2 k^2\over  M^2 + k^2} \ .
\end{equation}
The final answer is
\begin{equation}
{1\over e^2} \langle H_\mu (k)  H_\nu (-k)\rangle  =
\delta_{\mu\nu} - {k_\mu k_\nu \over M^2 +k^2} \ .
\end{equation}

Thus the correlation function of two gauge invariant operators $ {1\over 2}
\epsilon _{\alpha \beta \gamma} F_{\beta \gamma}
$ has only one longitudinal excitation as if one would calculate the two-point
correlator of the derivatives of a massive scalar field.

For the antisymmetric tensor field in $d=4$  things work not exactly the same
way but very close to it.
The only gauge invariant operator in our theory where we keep the antisymmetric
tensor coupled to the non-Abelian vector field is
\begin{equation}
H^\mu = -{i\over 2} \epsilon ^{\mu\nu\rho\sigma}\Bigl[ \partial _\nu
b_{\rho\sigma} +
{g^2
\over 32
\pi^2  }
A^a_\nu (F_{a \rho\sigma} - {g\over 3} f_{abc} A^b_\rho A^c_\sigma) \Bigr]
\equiv
h^\mu -i  \Omega^\mu \ .
\end{equation}
This is the only operator which is gauge invariant under both Maxwell and
Nicolai-Townsend non-Abelian symmetry. Separately, the operator $h^\mu$ which
is dual to a field strength of the antisymmetric tensor field  is invariant
under the Abelian gauge symmetry, but it is not invariant under the Yang-Mills
symmetry without the Chern-Simons term.

To examine our problem we may ignore small quantum   fluctuations of  the
radial component of the scalar field. These fluctuations are irrelevant when
one investigates the possibility that gauge symmetries protect axion theory
from getting a mass. Thus, the action we consider is the action (\ref{dual})
with the fixed radial component, $f(x)= f_0$,
\begin{equation}
{\cal L}_{b_{\mu\nu}}= {1\over 2}  f  _0^{-2}  H_{\mu\nu\lambda}^2 \ .
\label{dual2}\end{equation}

We may start by treating  the coupling of the $b_{\mu\nu}$-field to the
non-Abelian field perturbatively, i.e. we may study the correlator of two gauge
invariant operators $H_\mu$ by treating the coupling as a correction to the
value of the correlator of two operators $h_\mu$ without coupling.
The bare correlation function of two operators $h_{\mu}$ must be transverse
since
\begin{equation}
\partial _\mu h^\mu = -{i\over 2} \partial_\mu \epsilon ^{\mu\nu\rho\sigma}
\partial _\nu b_{\rho\sigma} = 0 \ .
\label{Bian}\end{equation}
Thus, as in the Polyakov case, we may expect
\begin{equation}
f_0^{-2} \langle H_\mu (k)  H_\nu (-k)\rangle  =  (\delta_{\mu\nu} - {k_\mu
k_\nu \over k^2} )
 + {k_\mu k_\nu \over k^4} \langle \rho(k) \rho(-k)\rangle  + ... \ ,
\end{equation}
where the dots correspond to the corrections to a transverse part of the
correlator.   We have parametrized the longitudinal excitation by some 2-point
correlator  of a ``charge density'' $\rho(x)$.  The divergence of our gauge
invariant operator  $H_\mu$, which we may associate with $\rho$, is given by
\begin{equation}
 f_0^{-1} \partial_\mu H^\mu (x) =  -i  f_0^{-1}  \partial_\mu  \Omega^\mu = -i
 {g^2
\over 32
\pi^2  f_0}   F_{\mu\nu}^a \, \tilde F^{a \mu\nu} (x)\equiv  -i \rho(x) \ .
\end{equation}

The correlator  of two divergences of the Yang-Mills Chern-Simons currents was
calculated by Shifman, Vainstein and Zakharov in ref.  \cite{invisible}. They
have shown that in a certain approximation it is proportional  to the axion
mass
\begin{equation}
\langle [ {g^2
\over 32
\pi^2  f_0 }   F_{\mu\nu}^a \, \tilde F^{a \mu\nu}] (k) \;  [ {g^2
\over 32
\pi^2  f_0 }   F_{\lambda\delta}^b \; \tilde F^{b \lambda\delta}] (-k)\rangle
\sim
 m_a^2 \ .
 \end{equation}
This serves  as an  indication that  one may obtain in the $b_{\mu\nu}$-theory
the longitudinal excitation of the gauge invariant operator corresponding to
a scalar massive particle instead of a massless $b_{\mu\nu}$\,-field.

One can  confirm these expectations by calculating the correlator  $\langle
H_\mu (k)  H_\nu (-k)\rangle$ directly,  without  separating our operator
$H_\mu$ into its free part $h_\mu$ and the interaction part $\Omega_\mu$. For
this purpose we
 will perform the generalized duality transformation in the functional integral
describing the theory.
Consider the ``first order"  functional integral for the action we have
considered above
with the additional term including the source $\eta^\mu$ to  the current
$f_0^{-1} H_\mu$:
\begin{equation}
Z(\eta) = \int {\cal D} a  {\cal D} H_\mu {\cal D} A_\alpha^a\  \exp \left[i
\int \Bigl(i f_0^{-1} \partial_\mu a \,  (H^\mu + i  \Omega^\mu) + {1\over 2}
f_0^{-2}
 H_\mu H^\mu + \eta^\mu  f_0^{-1} H_\mu + {\cal L}_{YM} \Bigr)\right] \ .
\end{equation}
The functional integration is performed over the pseudoscalar $a$ as well as
over the pseudovector $H_\mu$ and over the Yang-Mills fields.
For the sake of simplicity we do not write down explicitly the integration over
the ghosts  related to the gauge fixing of the non-Abelian symmetry.
We may evaluate this functional integral by integrating over the pseudoscalar
$a$ first. This will produce  a constraint given in eq. (\ref{con}). In this
way we will get rid of the $a$-integration
and the remaining integral will reduce to the integration over the constrained
$H_\mu $. Equivalently the integration over the constrained $H_\mu$  may be
replaced by the integration over the unconstrained $b_{\mu\nu}$  together with
the proper gauge-fixing procedure in the path integral. As before,    $H_\mu
(b, A) $ is defined in terms of
$b_{\mu\nu}$ and Yang-Mills fields  in eq. (\ref{con}). Thus we get
\begin{equation}
Z(\eta)  = \int {\cal D} b_{\mu\nu} D C_{\rm gh}  {\cal D} A_\alpha^a\ \exp
\left[ i \int{1\over 2}  f_0^{-2} H_\mu  H^\mu (b, A)
 + \eta_\mu f_0^{-1}  H^\mu (b, A)+ {\cal L}_{YM} + {\cal L}_{\rm gfix} +
{\cal L}_{\rm gh}\right] \ ,
\end{equation}
where we have written down the integration  over the complete set of ghost
fields $C_{\rm gh}$ related to both types of gauge symmetries.   ${\cal L}_{\rm
gfix}$ is the gauge-fixing part of the action and  ${\cal L}_{\rm gh}$ is the
action of the ghost fields.
By varying this functional twice over $\eta^\mu $ we will get
the correlations function of two gauge invariant operators $f_0^{-1} H_\mu (b,
A)$ defined in eq. (\ref{con}) and calculated in the dual version of the axion
theory.

On the other hand, we can perform the integration over $H_\mu$ first. This
does not change the fact that the second derivative of
$Z(\eta)$ gives the correlator of two $f_0^{-1} H$'s.  It is just an
alternative method of calculations. After integration over  $H_\mu$ we get
\begin{equation}
Z(\eta) = \int {\cal D} a   {\cal D} A_\alpha^a\  \exp \left[i \int  - {1\over
2}  (i \partial_\mu a  + \eta_\mu)^2  + {g^2 a \over   32 \pi^2 f_0} \;
F_{\mu\nu}^a \, \tilde F^{a \mu\nu}  + {\cal L}_{YM}  \right] \ .
\end{equation}
 The integration over the instanton configurations  in the dilute gas
approximation produces the effective action for the field $a$
\begin{equation}
Z(\eta) \approx  \int {\cal D} a   {\cal D} A_\alpha^a\  \exp \left[i \int
{1\over 2}  (\partial _\mu a  -i \eta_\mu )^2   + \Lambda^4_{QCD} \cos {a
\over f_0} + {\cal L}_{YM} \right] \ .
\end{equation}
Double variation of this generating functional gives us an expression for the
correlation function of two operators $f_0^{-1} H_\mu$ in the following form:
\begin{equation}
f_0^{-2} \langle H_\mu (k)  H_\nu (-k)\rangle =
\delta_{\mu\nu} - {k_\mu k_\nu \over m^2_a +k^2} \ .
\label{cor}
\end{equation}
This result is in a complete correspondence with the Polyakov result for $d=3$.
The first term in the r.h.s. of eq. (\ref{cor}) comes from the $\eta_\mu^2$
term in the generating functional, and the second term comes effectively from
the correlator of two derivatives of the massive field $a$.

Our conclusion is the following. The antisymmetric gauge field is coupled to
the Yang-Mills fields. With an account taken of instantons the behavior of the
system
becomes that of the massive scalar field theory. Note that this is {\it not} a
mass of the field $b_{\mu\nu}$. Rather it is the mass of the only
gauge-invariant degree of freedom associated with $b_{\mu\nu}$.

Thus we have  found a mechanism of getting small QCD mass for the axion in the
dual version of the theory, which does not violate any of the gauge symmetries.
This result has an important implication  that there is no reason to expect
that the gauge symmetry which exists in the dual formulation of the axion
theory can protect the axion mass from becoming very large because of the
gravitational   effects.

\

\section{\label{3s} Axionic instantons, or wormhole solutions with  fixed
radial component of the scalar field}

The axionic  wormholes  which may provide mass to the axion via
gravitational effects have been found originally in the dual version of the
axion theory with antisymmetric tensor field \cite{GS}. The theory of the
complex scalar field does not have such wormhole solutions  unless the
functional integral is supplemented by the proper boundary conditions.  The
corresponding investigation has been performed in \cite{Lee} and in \cite{CL}.
The conclusion was that there exists a consistent procedure  to obtain the same
wormhole solutions in both versions of the axion theory.

In this section we discuss wormholes with a frozen radial component of the
field, like Giddings
and Strominger \cite{GS}  and  Lee \cite{Lee}. We will mainly reproduce their
results. However, in addition we will discuss the subtlety related to the
 boundary terms and the value of the Euclidean action. We will find out
that  if one considers the configuration with the  boundary at the wormhole
throat, see Fig. 1 (a),
and calculates the contribution to the action from the boundary terms in a
standard way, one
ends up with the action which is only $(1- {2\over \pi}) \sim 0.36$ of the
original
action, which was calculated in \cite{GS,Lee} without an account taken of the
boundary
terms on the wormhole throat.\footnote{We are grateful
to A. Strominger for the discussion of this subtlety.}

Giddings and Strominger have found their wormhole solution in the
string-inspired theory with the action
 \begin{equation}\label{String}
S_{\rm H} = \int d^4 x \sqrt g \left( - {M_{\rm P}^2 \over 16 \pi } {\cal R} +
f_0^{-2}
H_{\mu\nu\lambda}^2
\right)- {M_{\rm P}^2 \over 16 \pi } \int_{\partial V} d^3 S (K - K_{0}) \ .
\end{equation}
 Here $H_{\mu\nu\lambda} = \partial _{[\mu} b_{\nu\lambda]}$ is the field
strength of the antisymmetric tensor field.
The last term in the action  is the Gibbons-Hawking surface term.

Meanwhile, Lee obtained wormhole solutions in the theory of a pseudoscalar
axion field $\theta \equiv {a\over f_0}$ with the  action
\begin{equation}\label{LEE}
S = \int d^4 x \sqrt g \left( - {M_{\rm P}^2 \over 16 \pi } {\cal R} + {1\over
2}f_0^{2}\,
(\partial_{\mu} \theta)^2
\right)- {M_{\rm P}^2 \over 16 \pi } \int_{\partial V} d^3 S (K - K_{0}) \ .
\end{equation}

The wormhole geometry has the following form
\begin{equation}
ds^2 = dr^2 + R(r)^2 d^2 \Omega_3 \ .
\label{wh}
\end{equation}

As we already mentioned,   there are no true solutions of
the Lagrange equations following from (\ref{LEE}). However,  it was pointed out
in \cite{Lee,CL} that these solutions appear if one   takes into account the
charge conservation condition in  space (\ref{wh}). The global charge $n$,
defined as an integral over the 3-space from the zero component of the Noether
current,
\begin{equation}\label{charge}
n =  \int_{S_{3}} R^3 f^2 {\theta}' = 2\pi^2 R^3 f_0^2 \theta' = const \ ,
\end{equation}
is an integer upon quantization.\footnote{Note that in the situation with
spontaneous symmetry the global charge is carried not by charged particles, but
by the vortices of the classical scalar field with the time-dependent phase
$\theta$.}

The conclusion of ref. \cite{Lee,CL} was that if one makes the variation of the
action under the condition $n = const$, one obtains equations of motion which
are equivalent to the equations in the theory (\ref{String}). In particular,
  the gravitational equation of motion in both theories looks like
\begin{equation}
R'(r)^2=1-\left( {R(0) \over  R(r)} \right)^4 \ .
\end{equation}
Here  $R(0)$ is the size of the throat of the wormhole defined by the condition
$R'(0) =0$. It is    given by
\begin{equation}
R(0)= \left({n^2 \over  3 M_{\rm P}^2 \pi^3 f_0^2 }\right)^{1\over 4} \ .
\end{equation}
 The wormhole geometry $R(r)$ can be expressed
analytically  in terms of  elliptic integrals \cite{GS}.

The   wormhole action with account taken of the boundary terms both
at the outer boundary at $r=r_0,  \; r_0 \rightarrow \infty$, and at the inner
boundary
at $ r=0$  is
\begin{equation}
S_{\rm total} = {3 M_{\rm P}^2\over 4 \pi }\int d\Omega \int_0^\infty
dr  R(r) R'(r) (1- R'(r) ) \ .
\end{equation}
The integration can be performed as follows
\begin{equation}\label{total}
S_{\rm total}  = {3 M_{\rm P}^2\over 4 \pi } \; 2\pi^2 \;{1\over 2}
\int_{R^2(0)}^{\infty}  d  R^2
\left(1 - \sqrt{1-  {R^4(0) \over  R^4}}\,\right)= {3\pi^2
\over 8} M_{\rm P}^2  R^2 (0)\cdot \left(1- {2\over \pi}\right) \ .
\end{equation}

Giddings-Strominger-Lee on shell action (without the inner boundary terms) is
\cite{GS}
\begin{equation}\label{nobound}
S_{\rm no bound} = {3\pi M_{\rm P}^2\over 2} \int_0^\infty  dr  R(r)  (1-
[R'(r)]^2 )= {3\pi^2
\over 8} M_{\rm P}^2  R^2 (0) \ .
\end{equation}
The boundary term which we added,
\begin{equation}\label{surf}
S_{\rm bound} = - {3\pi M_{\rm P}^2\over 2} \int_0^\infty  dr  R(r)  (1-
R'(r))= {3\pi^2
M_{\rm P}^2  R^2 (0)\over 8} \cdot \left(- {2\over \pi}\right) \ ,
\end{equation}
is $- 0.637$ of the action without the boundary term.  To understand
better the boundary term contribution consider it in the form
\begin{equation}
S_{\rm bound}=  {3 M_{\rm P}^2 \over 8 \pi} \int d\Omega \int _0^\infty dr
{\partial\over
\partial r} \Bigl[R^2(r) (1- R'(r) )\Bigr] \ .
\label{neck}\end{equation}
The contribution comes only from the throat where $R(r) = R(0)$, $ R'(r) =0$.
It
equals
\begin{equation}
S_{\rm bound}=  -{3 M_{\rm P}^2 \over 8 \pi}\,  2\pi^2\, R^2(0) = -{3  \pi
\over 4}
M_{\rm P}^2 R^2(0) \ ,
\end{equation}
which is in a complete agreement with (\ref{surf}). Note that the extrinsic
curvature term $K$ at the $S^3$ boundary at $r=0$  (the second term in eq.
(\ref{neck}))  vanishes, in agreement with the discussion in \cite{STRBOOK},
but an additional
non-vanishing contribution comes from the term
$K_0$, which was added to remove the divergence of the action at the outer
boundary.

The total action for the configuration
with the boundary at the throat is thus only $0.363$ of the value of the action
without the surface term. This gives an additional support to the idea that in
gravitational problems one has to be very careful with the boundary terms. The
outer boundary surface term $K$ has to be corrected by $K_0$,
otherwise
the action is infinite. Does it mean that the surface term has to have the
same functional form $K-K_0$ on both boundaries? If the answer is yes, we have
to subtract ${2\over \pi}$ part of the action and make it almost three times
smaller than the action obtained in \cite{GS,Lee}.

One should note,   that the problem of obtaining a proper contribution from the
inner boundary is very complicated. For example, recently it was argued that an
adequate account taken of the inner boundary of an extreme charged black hole
changes its Euler number \cite{GK}, which makes its entropy zero
\cite{Hawking,Teit,GK}. However, even for the black hole case this issue is
rather nontrivial. The situation with the wormholes is even more complicated
and ambiguous. If  one considers the
wormholes without the inner boundary contribution as in \cite{GS,Lee,AW,CL},
one has to add to our result for the action the universal term
${3  \pi \over 4}  M_{\rm P}^2 R^2(0)$, which depends only on the size of the
wormhole throat.  One may also want to calculate the action on a symmetric
configuration $-\infty <r < +\infty$, Fig. 1 (b).  In this case   one will not
have any   contribution from the inner boundary, and the action will be two
times greater than the GSL action (\ref{nobound}).

In what follows we will work with the action defined with both boundaries and
with the surface term $K-K_0$ on both boundaries since this prescription seems
to be more internally consistent.  Another advantage of this prescription is
that it gives us the smallest   action as compared with other prescriptions
mentioned above. Therefore if we find a way to avoid the strong violation of
global symmetries in our approach, we will simultaneously solve the
corresponding problem in other approaches as well.

The radius of the    throat  of the Giddings-Strominger-Lee wormhole depends on
the parameter $f_0$ and
on the value of the charge $n$. We may therefore express the action in terms of
these
parameters as follows:
\begin{equation}\label{frozen}
S_{\rm total} = {\sqrt{3 \pi} \over 8 } \,  \Bigl(1- {2\over \pi}\Bigr)
 {nM_{\rm P} \over      f_0 } \ .
\end{equation}
For completeness, we will give here also the value  which has been obtained in
\cite{GS,Lee} without an account taken of the inner boundary,
 \begin{equation}
S_{\rm no bound} = {\sqrt{3 \pi} \over 8 }  \left({n M_{\rm P} \over      f_0
}\right) \ .
\end{equation}

If one takes the smallest action (\ref{frozen}) with $ n = 1$ and $f_0 =
10^{12}$ GeV, one obtains an enormously strong suppression $\sim
\exp\bigl(-10^{6})$. This would immediately solve the problem of the global
symmetry violation. Unfortunately, however, things are much more complicated.
As it was pointed out in \cite{AW}, it is almost impossible to keep the field
$f(x)$ close to $f_0$ on the wormhole solution. In what follows we will show
that if one allows the field $f$ to depend on $r$,   this field typically grows
to $f\sim M_{\rm P} \gg f_0$ near the wormhole throat. Therefore one
needs to make a separate investigation to calculate the wormhole action
in realistic
theories with spontaneous symmetry breaking. This investigation will be
contained in the next section.

\

\section{\label{4s}  Wormhole solutions with  dynamical complex scalar}
\vskip -.6 cm
{\Large\bf  ~~~~~~field
  and spontaneous symmetry breaking}
\vskip 0.2 cm

\subsection{\label{4.1} Equations for the scalar field in the wormhole
geometry}
In this section we will  discuss wormhole solutions  for several different
theories with  spontaneous symmetry
breaking. In particular, we will  investigate the dependence of the action on
the value
of the vacuum expectation value $f_0$ of the radial part of the scalar field
far away
from the wormhole.

We will   study   interaction of gravity with the complex scalar field
$\Phi(x) =
{ f(x)\over \sqrt 2}\, e^{i \theta(x)}$.
The corresponding action is
\begin{equation}
S = \int d^4 x \sqrt g \left( - {M_{\rm P}^2 \over 16 \pi } {\cal R} + {1\over
2}
(\partial_{\mu} f)^2 +
{1\over 2}f^{2}
(\partial_{\mu} \theta)^2 + V(f)
\right)- {M_{\rm P}^2 \over 16 \pi } \int_{\partial V} d^3 S^a (K_a - K_{0a}) \
{}.
\end{equation}

We will assume that the potential $V(f)$ has a   minimum at some
value
$f= f_0$.  The vacuum energy vanishes in this minimum,
\begin{equation}
V'(f)|_{f=f_0} = V(f)|_{f=f_0} = 0 \ .
\end{equation}

Equations of motion corresponding to the analytic
continuation of the Euclidean theory of the 2-form version of the axion theory
\cite{Lee,AW,CL} are
\begin{eqnarray}\label{EQ}
f'' -{3R'f' \over R} - {dV(f) \over df}- {n^2 \over 4 \pi^4 f^3 R^6}&= &0 \ ,
\\
\label{MAINEQQ}
\nonumber\\
R'^2- 1+ {8\pi \over 3M_{\rm P}^2} R^2\left( V(f)+{n^2 \over
8\pi^4 f^2 R^6}-{f'^2\over
2}\right) &=& 0 \ .
\end{eqnarray}
and the value of ${\theta}' = {n\over 2\pi^2 f^2  R^3}$ has been already
substituted
in the equations.

Using these equations, one can derive the following expression for the wormhole
action:

\begin{equation}\label{ACTION}
S_{\rm total} = 2\pi^2  \int_0^\infty  dr \left(R^3    f'^2
  + {3 M_{\rm P}^2\over 4 \pi } R R' (1- R')\right) \ .
\end{equation}

We are looking for the wormhole solutions with the geometry given in eq.
(\ref{wh})
and with the fields $R(r)$ and  $f(r)$   solving the system of equations
(\ref{EQ}), (\ref{MAINEQQ}).  Our
boundary conditions are
\begin{eqnarray}\label{COND1}
R'(0)&=&0\ , \\
\label{COND2} f'(0)&=&0\ , \\
\label{COND3} f'(r)&\rightarrow&0\ , \  \  r\rightarrow \infty\ ,\\
\label{COND4} f(r) &\rightarrow&f_0\ , \  \ r\rightarrow \infty\ .
\end{eqnarray}
We will examine the following potentials:
\begin{eqnarray}
\label{POT1} V_1(f)&=&\frac{\lambda}{4}\left( f^2-f_0^2 \right)^2\ ,  \\
\label{POT2} V_2(f)&=&\frac{\lambda}{6M_{\rm P}^2} \left(f^6 - 3f_0^4 f^2 + 2
f_0^6 \right)\ ,  \\
\label{POT3} V_3 (f)&=&\frac{\lambda}{4}e^{\beta f M_{\rm P}^{-1}}\left(
f^2-f_0^2 \right)^2 \ .
\end{eqnarray}
The first potential is a standard potential of a theory with spontaneous
symmetry breaking, the second one is inspired by some phenomenological models
based on supergravity with $ f^2_0 \sim  { {M_{p}  m_{3/2}\over\sqrt \lambda}
}$, the third one is inspired by  string theory.
We will find the wormhole solutions numerically using the fact that the field
$f$
has a definite value $f_0$  at infinity.  In particular, for axions we will be
interested
in
$f_0 \sim 10^{12}$ GeV  and for the textures in $f_0 \sim 10^{16}$
GeV.\footnote{We should emphasize again  that the value of the parameter $f_0$
may be quite different from $10^{12}$ GeV. However, as we will see, our results
depend on $f_0$ only logarithmically. Also, for   textures one should take a
theory with another group of symmetries, which, however, should not
considerably change our results.} Typically the radial component of the axion
field $f$ will grow many orders of magnitude from the value $f_0$ to $f \sim
M_{\rm P}$ when approaching the wormhole throat.

Note that  our expression for the total action (\ref{ACTION}) does not have any
{\it explicit} dependence on the effective potential $V(f)$ and on the charge
$n$.
It is important also that $0\leq R'(r) < 1$ for all wormhole solutions which we
are going to study. As a result, the integrand is always positive, despite the
fact that the gravitational contribution to the action may be negative. One of
the consequences of this result is that if we make a cut-off at some radius
$r_c$ and integrate from $r = 0$ to $r = r_c$, the result will be always
smaller than the total action (\ref{ACTION}).

\subsection{\label{4.2} Wormholes in the theories without symmetry breaking}

In this paper we study the theories with spontaneous symmetry breaking.
However,   on a small scale $R {\
\lower-1.2pt\vbox{\hbox{\rlap{$<$}\lower5pt\vbox{\hbox{$\sim$}}}}\ }m^{-1}(f)
\sim ({\sqrt\lambda f})^{-1}$    the wormholes  in the theories with the
effective potential  ${\lambda\over 4} f^4$ described in   \cite{AW,CL}    are
very similar to the   wormholes which we have found in  many theories with
spontaneous symmetry breaking. Therefore we will briefly discuss here the
wormhole solutions in the theory ${\lambda\over 4} \phi^4$, following
\cite{AW}.

Near the wormhole throat the solution was obtained in \cite{AW} numerically.
This solution should be matched to the solution obtained analytically at large
$r$. According to eq. (\ref{MAINEQQ}), far away from the throat one has $R(r) =
r$. Therefore equation (\ref{EQ}) in this case (for n = 1) reads
\begin{equation}\label{AW1}
f'' -{3 f' \over r} - {\lambda f^3}- {1 \over 4 \pi^4 f^3 r^6} =  0 \ .
\end{equation}
This equation has an exact solution
\begin{equation}\label{AW2}
f = {\delta \over r} \ ,
\end{equation}
where $\delta $ is determined from the equation
\begin{equation}\label{AW3}
{1 \over 4 \pi^4 \delta^4}- \lambda\delta^2 = 1 \ .
\end{equation}
If one adds to the effective potential   the term ${m^2\over 2} f^2$
with $m^2 > 0$ (no spontaneous symmetry breaking), the asymptotic behavior at
large $r$ changes to $f \sim (2\pi^2 m r^3)^{-1/2}$.

It was assumed in \cite{AW} that the contribution to the action from the
region near the wormhole throat (until the solution approaches its regime
(\ref{AW2})) is very small.
On the other hand, integration far away from the throat gives the contribution
\begin{equation}\label{AW4}
S = 2\pi^2\int r^3 dr \left( {1\over 2} f'^2 + {1 \over 8 \pi^4 f^2 r^6} +
V(f)\right) \ .
\end{equation}

The total action integrated up to some $r_{\rm max}$ has the following general
form:
\begin{equation}\label{AW5}
S_{\rm total}(r_{\rm max}) = -(1 - {3\lambda\over 8\pi^2}) \ln (mr_n)+ {mr_{\rm
max}} + \Delta S \ .
\end{equation}
Here $r_n$ corresponds to the place where the numerical solution near the
wormhole throat matches the solution (\ref{AW2}), $\Delta S$ stands for several
other terms which have not been determined in \cite{AW}. One can  calculate
$\Delta S$ using our expression for the action (\ref{ACTION}); typically this
term is fairly small, $\Delta S = O(1)$.

Thus the two most interesting terms in (\ref{AW5}) is the logarithmic term and
the term which linearly diverges at large $r_{\rm max}$. As we will see, a
similar logarithmic contribution appears in the theories with spontaneous
symmetry breaking as well. However, the linear divergence at    $r_{\rm max}
\to \infty$ is a particular property of the theory without symmetry breaking,
which is a consequence of the asymptotic behavior $f \sim (2\pi^2 m
r^3)^{-1/2}$ at large $r$. One could conclude that the wormhole action is
infinite in the theories without spontaneous symmetry breaking, and therefore
these theories cannot lead to the global symmetry violation.

 However, it was argued in \cite{AW,CL} that  this is not the case, and in fact
the effects of global symmetry violation are quite significant. It was
suggested that these effects are suppressed only by some small part of action
$S_w$ coming from the region close to the throat of the wormhole. The value of
$S_w$ was not calculated in \cite{AW,CL}, but it was estimated to be  $O(1)$.
We will return to the discussion of this issue when we will consider wormholes
in the theories with spontaneous symmetry breaking.

\subsection{\label{4.3} Wormholes in the theory with the simplest potential
${\lambda\over 4}{(f^2 - f^2_0)^2}$}

Equation for the scalar field $f$ in the theory ${\lambda\over 4}{(f^2 -
f^2_0)^2}$ on the wormhole configuration looks very similar to eq. (\ref{AW1}):
\begin{equation}\label{SP1}
f'' -{3 f' \over r} - {\lambda f(f^2 -f_0^2)}- {1 \over 4 \pi^4 f^3 r^6} =  0 \
{}.
\end{equation}
Therefore   the wormhole solutions at $f \gg f_0$ behave just as their
counterparts in the theory without spontaneous symmetry breaking.  In
particular, far away from the wormhole throat
$f = {\delta \over r}$, where $\delta $ is determined by
${1 \over 4 \pi^4 \delta^4}- \lambda\delta ^2 = 1$.

The main difference between the wormhole solutions with and without symmetry
breaking appears on the scale $r {\
\lower-1.2pt\vbox{\hbox{\rlap{$>$}\lower5pt\vbox{\hbox{$\sim$}}}}\
}\Bigl(4\sqrt 2 \pi^3\lambda \Bigr)^{-1/5} f_0^{-1}$, where   the field $f$
approaches $f_0$ very rapidly (though not exponentially, as   anticipated in
\cite{AW}). We have found that at large $r$
\begin{equation}\label{larger}
f(r)-f_0 = {1\over 8\pi^4 \lambda f_0^5 r^6} \ ,
\end{equation}
which leads to a finite (and very small)  contribution to action from the
region $r {\ \lower-1.2pt\vbox{\hbox{\rlap{$>$}\lower5pt\vbox{\hbox{$\sim$}}}}\
} \Bigl(4\sqrt 2 \pi^3\lambda \Bigr)^{-1/5} f_0^{-1} $. Thus, in our theory we
do not have any problems with infinite wormhole action.

To study the behavior of our solutions in all regions from $r = 0$ to $
r\to\infty$, and to avoid having numerical uncertainties associated with the
value of action coming from each of the regions we performed a numerical
investigation of the wormhole solutions in our theory. This investigation was
rather complicated since the solutions happen to be extremely sensitive to the
boundary conditions. Whereas it was possible to obtain some results for $f_0
\sim 10^{18}$ GeV using standard numerical recipes \cite{PTVF}, it was
necessary to substantially improve  these methods in order to study the
physically interesting regime  with $f_0 \sim 10^{12}$ GeV. Since this
improvement may be useful not only for finding the wormhole solutions, we will
describe our method in the Appendix.

We have found it useful   to make our numerical calculations in  dimensionless
variables $\rho$, $A(\rho)$, $F$ and $U(F)$, where
\begin{equation}
  \rho  = r M_{\rm P} \; \sqrt {  3\lambda \over 8\pi}\ , \qquad A = R M_{\rm
P} \; \sqrt {  3\lambda \over 8\pi}\ , \qquad F= {f\over M_{\rm P}} \sqrt {8\pi
  \over 3}\ ,\qquad U(F) \equiv  \lambda^{-1} V(\phi)  \ .
\end{equation}
We also introduce the  combination $Q\equiv {n^2 \lambda^2 \over 8\pi^4}$
\cite{AW}.

Equations of motion   in these variables
are given by
\begin{eqnarray}\label{MAINEQ}
F''(\rho) +{3A'(\rho)F'(\rho) \over A(\rho)} - {dU(F) \over dF}-
{2 Q^2 \over
F^3(\rho) A^6 (\rho)}&= &0 \ , \\
\label{MAINEQ2}
\nonumber\\
 A'^2(\rho) - 1+  A(\rho)^2\left( U(F)+{Q^2 \over
 F^2 (\rho) A^6 (\rho)}-{1
\over
2} F'^2(\rho) \right) &=& 0 \ .
\end{eqnarray}

The on shell action is
\begin{equation}
S_{\rm total} = {n\over \sqrt {2Q} } \int_0^\infty  d\rho \left[
A^3(\rho)F'^2(\rho)  +
  {2  A(\rho) A'(\rho) (1- A'(\rho)} \right] \ .
\end{equation}

Our numerical solutions of the system of differential eqs. (\ref{MAINEQ})
for the theory ${\lambda\over 4}{(f^2 - f^2_0)^2}$ depend
on the value of the coupling constant $\lambda$ and on the asymptotic
value of the field $f_0$. In all cases we are interested in the strongest
 possible violation of global symmetries by gravity and therefore we will
consider
the smallest value of the charge $n=1$. We will present most of our results
for $\lambda = 0.1$  and
$\lambda=1$, but in all figures we will show only the case $\lambda = 0.1$.
Different asymptotic values of the field $f_0$ are considered,
  from $f_0= 10^{12}$ GeV  to $f_0= 10^{18}$ GeV. The corresponding values of
the  dimensionless field $F_0$ are
\begin{equation}
F_0 = f_0 \sqrt {8\pi \over 3 M_{\rm P}^2} =  f_0\times 2.4 \times10^{-19}\,
{\mbox GeV}^{-1}  \   .
\end{equation}
For example for $f_0= 10^{12}$ GeV one has $F_0 = 2.4\times 10^{-7}$,
and for  $f_0= 10^{16}$ GeV one has $F_0 = 2.4\times 10^{-3}$.
The solution for the dimensionless radial component of the scalar field $F $ is
represented on Fig. 2 where the value of $\log_{10} F $ is plotted as a
function of
$\log_{10} \rho $ for different seven values of $F_0$, corresponding
to $f_0 = (10^{18}, 10^{17}, 10^{16}, 10^{15}, 10^{14}, 10^{13}, 10^{12})\, \,
{\mbox GeV}$.
One can see from Fig. 2    that
 all solutions with different asymptotic values of $f_0$ behave in the same way
near to the wormhole throat. This means that the
axion type field
$F$ which far away from the  wormhole was $\sim 10^{-7}$, increases seven
orders of
magnitude  near the throat to become $\sim 1$, and the texture-type field $F$
which far away from the  wormhole was $\sim 10^{-3}$, increases three orders of
magnitude  near the throat to reach the same value $\sim 1$ corresponding to $f
\sim M_{\rm P}$.
Thus, the solutions which we obtained differ very much from the
Giddings-Strominger-Lee solutions with a fixed value $f(r) = f_0$. On the other
hand, the fact that the solutions at small $r$ do not depend on $f_0$ confirm
our expectations that the behavior of these solutions near the wormhole throat
does not depend on spontaneous symmetry breaking.

This conclusion becomes even more obvious if one consider
Fig. 3, which gives the value of the function $A(\rho)$ (i.e. $R(r)$). This
function, describing the
geometry is completely insensitive to the asymptotic value of the field $F_0$
(i.e. of $f_0$).
Therefore we have only one curve for all  seven cases above.

One can also express  our results in the usual dimensional variables. One can
show, in particular, that if the coupling constant  $\lambda $  is very small,
then,  just as in the theory without symmetry breaking \cite{AW}, the value of
the field $f$ at the wormhole throat and the radius of the throat are given
by the simple expressions independent on $\lambda$ and $f_0$:
\begin{equation}\label{center}
f(0) \approx M_{\rm P} \sqrt{3\over 8\pi} \ , ~~~~ R(0) \approx {2 \over
\sqrt{3\sqrt 2\pi}} M_{\rm P}^{-1} \ .
\end{equation}

Meanwhile the total action does depend on $f_0$, and this dependence is pretty
simple. Fig. 4 represents the value of the action as the function of $-\ln F_0$
for
$\lambda=0.1$. The black line and dots give the total action with the boundary
term,
the grey one shows the action without the contribution of the boundary term at
the inner boundary (see the discussion of this possibility in the previous
section). All data fit  the following simple equation for the action
\begin{equation}\label{ln}
S_{\rm total}=a-b\ln F_0 \ .
\end{equation}
The values of  $a$, $b$  for two values of $\lambda$ are
\begin{eqnarray}
a&=&0.186\pm 0.005\ ,  \qquad b=1.001\pm 0.005\ ,  \qquad \lambda=0.1 \ , \\
\nonumber\\
a&=&0.188\pm 0.01\ , \qquad b=1.010\pm 0.002\ ,  \qquad \lambda=1 \ .
\end{eqnarray}
Thus the dependence on $\lambda$ is not strong and the total action is small.

If we consider the configuration without the boundary term, the action
increases slightly.
The values of $a$ and $b$ for two values of $\lambda$ are
\begin{eqnarray}
a&=&0.850\pm 0.005,  \qquad b=1.001\pm 0.005 \qquad \lambda=0.1 \ ,\\
\nonumber\\
a&=&1.33\pm 0.01, \qquad b=1.010\pm 0.002 \qquad \lambda=1 \ .
\end{eqnarray}
However, even in this case the total action remains of the order of $15$ for
$f_0 \sim 10^{12}$ GeV. This is much smaller than what we need.

This conclusion may seem somewhat unexpected. Indeed, the only natural length
scale in the theory of gravity is $M_{\rm P}^{-1}$. How could it be possible
for a wormhole with a throat of a radius $\sim M_{\rm P}^{-1}$ to eat a vortex
with $f^2\dot \theta \not = 0$ or a particle of a typical size $m^{-1} \gg
M_{\rm P}^{-1}$? Indeed, if   wormholes do not change the value of the scalar
field $f$, such processes are extremely strongly suppressed, as we have seen
for the case of the Giddings-Strominger-Lee wormhole. However, in our case the
total action is rather small, and it depends on $f_0$ only logarithmically. A
possible interpretation of our results is the following. It does not cost the
wormhole almost any action to squeeze the vortex to the size $r \sim
\lambda^{-1/5} f_0^{-1}$. Later (in the Euclidean time $r$) by increasing the
scalar field $f$ the wormhole squeezes the vortex to the Planckian size and
easily swallows it. One may say that our wormholes  have a small throat but a
very big mouth; they compactify the charge before eating it.

Note  that the total action   provides the maximum value of  suppression of
the violation of global symmetries. However, as we have already mentioned, this
suppression in fact may be even much weaker, if it is controlled not by the
total
action but only by the contribution to the action from the small vicinity of
the wormhole throat \cite{AW,CL}.

Indeed, nonperturbative effects are controlled by the total action only if one
can use the dilute gas approximation and consider contribution of each wormhole
separately. If the wormholes are very compact and their action is very large,
then it is indeed the case. Otherwise one may consider a possibility that the
wormholes carrying away opposite charges  can screen the large-scale ``tails''
of each other, and their effective action then will be determined by the
integration from $r = 0$  to $r_{c}$, where $2r_{c}$ is a typical distance
between the throats of different wormholes. If $r_c$ is not much different from
the radius of the wormhole throat $R(0)$, the corresponding action should be
small, and there will be no suppression of the wormhole-induced global symmetry
violation.

To make an estimate of the cut-off radius $r_{c}$ (assuming that we are already
in the regime $R \approx r$) one may write an approximate condition implying
that there are no wormholes within the distance $2r_c$ from each other:
\begin{equation}\label{screen}
2\pi^2 \Bigl({r_c\over R(0)}\Bigr)^4 \, e^{- S_{\rm total}(r_c)} \sim 1 \ .
\end{equation}
Here $e^{- S_{\rm total}(r_c)}$ appears due to the exponential suppression of
the wormhole-like fluctuations on the scale $r_c$, and $2\pi^2 \Bigl({r_c\over
R(0)}\Bigr)^4$ is our estimate for the subexponential factor.

Using these results and eq. (\ref{screen}) one can obtain the value of $e^{-
S_{\rm total}(r_c)}$ which should be associated with the effective coupling
constant of the operators violating global symmetries. In the case we are
considering right now it is a pretty easy problem to solve. Indeed, at $r \gg
R(0)$ our wormhole solution enters the regime $R = r$, $f = {\beta\over r}$,
and its action at this stage with a very good accuracy is equal to $\ln {r\over
R(0)}$ (compare with (\ref{ln})). Therefore our condition (\ref{screen}) in
this case reads
\begin{equation}\label{screen2}
2\pi^2 \Bigl({r_c\over R(0)}\Bigr)^4 \, e^{- \ln {r\over R(0)}} = 2\pi^2
\Bigl({r_c\over R(0)}\Bigr)^3 \sim 1 \ .
\end{equation}
It is clear that this condition cannot be satisfied for $r > R(0)$. Thus, $r_c
{\ \lower-1.2pt\vbox{\hbox{\rlap{$<$}\lower5pt\vbox{\hbox{$\sim$}}}}\ }R(0)$
for the wormholes in the theory ${\lambda\over 4}{(f^2 - f^2_0)^2}$. This gives
$S_{\rm total}(r_c) = O(1)$, which implies existence of  the vertex operators
$g M_{\rm P}^3(\Phi +\Phi^*)$ violating global U(1)  symmetry with the
unacceptably large effective coupling constant $g \sim e^{-S_{\rm total}(r_c))}
= O(1)$.

Thus, spontaneous symmetry breaking {\it per se} does not imply any suppression
of the wormhole-induced violation of the global symmetries. One should be
warned, however, that this conclusion was based on a rather crude estimate of
$e^{-S_{\rm total}(r_c)}$ using eq. (\ref{screen}). This equation is based on
the assumption that if one has many wormholes at a distance $2 r_c$ from each
other, the action of each wormhole is (approximately) equal to the action of a
single wormhole solution integrated from $r = 0 $ up to $r_c$.  Note, however,
that the scalar field $f$ at $r = r_c$ remains extremely large, $f(r_c) \sim
M_{\rm P} \gg f_0$. Thus one could argue  that if  space were populated by many
wormholes displaced at a very small distance from each other, this would not
describe our original situation where the average amplitude of the radial
component is equal to $f_0 \ll M_{\rm P}$. This argument does not really
invalidate the multi-wormhole scenario. The radial part of the field $\Phi$ may
be quite large in the vicinity of each wormhole, but the presence of the charge
implies that the phase $\theta$ depends on $r$ and can be different for each of
the wormholes. Therefore, even  though the  value of the field $\Phi$ near each
of the wormholes is of the order of $M_{\rm P}$, the  average field $\Phi$ in
the whole space (after taking the average over fluctuations of $\theta$) can
have a very small radial component $f_0 \ll M_{\rm P}$.

To return to a more solid ground, one should note that even if one does not
want to consider this multi-wormhole picture and decrease the wormhole action
by introducing a cut-off at $r = r_c$, one still has a problem. Indeed, in the
theory with spontaneous symmetry breaking, unlike in the theory without
symmetry breaking considered in \cite{AW,CL}, the total wormhole action is
finite {\it and small}. The largest action which we have obtained for the axion
theory with $f_0 \sim 10^{12}$ GeV is only about 15. This   is more than ten
times smaller than the action $S \sim 190$ which is necessary to save the axion
theory. Therefore we will study now other, more complicated models, where one
may hope to obtain larger values of $S$.

Note that if we find the wormholes which have a very large action $S {\
\lower-1.2pt\vbox{\hbox{\rlap{$>$}\lower5pt\vbox{\hbox{$\sim$}}}}\ }190$ given
by  the integration in a small vicinity of the wormhole throat (and we will
find such solutions), then our equation (\ref{screen}) will suggest that the
cut-off radius $r_c$ should be many orders of magnitude greater than $R(0)$.
Therefore in all situations where we will find a solution to the problem of
strong violation of the global symmetries, the effective coupling constants in
our vertex operators of the type of $g_1 M_{\rm P}^3(\Phi +\Phi^*)$ will be
determined by the total wormhole action $S_{total}$ rather than by its small
part originated by the integration near the wormhole throat. This will
eliminate all uncertainties with the interpretation of the vertex operators and
multi-wormhole solutions described above.

\subsection{\label{4.4} A more complicated polynomial potential}
The next theory on our list has an effective potential
\begin{equation}\label{six}
V_2(f) = \frac{\lambda}{6M_{\rm P}^2} \left(f^6 - 3f_0^4 f^2 + 2 f_0^6 \right)
\ ,
\end{equation}
 where $ f^2_0 \sim  { {M_{p}\,  m_{3/2}\over\sqrt \lambda} }$.
This potential  was suggested to us by M. Dine as a useful
phenomenological potential for the axion theory  which might follow from
supergravity. We have found the wormhole solution
in this theory and performed the calculation of the action. There was
practically no difference in the value of the action as compared with the
action in the theory   ${\lambda\over 4}{(f^2 - f^2_0)^2}$. Again we have
recovered the logarithmic dependence on the value of $f_0$.
All figures practically coincide with the ones obtained in the previous case.
The conclusion is
that the mild change of the potential  responsible for spontaneous symmetry
breaking does not change the value of the wormhole action and therefore in
such theories we have to face a strong violation of global symmetries.

The reason can be easily understood. In the theory ${\lambda\over 4}{(f^2 -
f^2_0)^2}$ the  field $f $ far away from the wormhole throat  behaved as $f =
{\delta \over r}$, with $\delta $ being determined by  ${1 \over 4 \pi^4 \delta
^4}- \lambda\delta^2 = 1$. When the field $f$ decreases as  $f = {\delta \over
r}$, the  contribution of the term $\lambda f^3$ to the equation of motion
(\ref{SP1}) decreases as fast as the contribution of all other terms (for $f
\gg f_0$).   If now one has an effective potential which depends on $f$ as
$f^6$, then its contribution to the equations of motion in the regime $f \sim
{\delta \over r}$   decreases even faster than other terms, and the field
behaves as in the theory ${\lambda\over 4}{(f^2 - f^2_0)^2}$ in the small
$\lambda$ limit.

Thus the existence of the regime $f \sim {\delta\over r}$ is a very general
property of  our wormhole solutions. This leads to the familiar logarithmic
dependence of the action on $f_0$. Therefore it is very difficult to increase
the action $S$ by changing $f_0$. However, as we will see now, one can
considerably increase the   action   if one succeeds to increase the radius of
the wormhole throat   $R(0)$.

\subsection{\label{4.5} Exponential potential}

 The third class of potentials  includes an
exponential dependence on  the radial component of the field which forces
the field to remain close to its minimum value and not grow so fast as in the
previous cases, $V_3 (f) =\frac{\lambda}{4}e^{\beta f M_{\rm P}^{-1}}\left(
f^2-f_0^2 \right)^2$. In terms of our dimensionless variables the potential is
\begin{equation}
U_3 (F)=\frac{1}{4}e^{\beta F}\left( F^2-F_0^2 \right)^2 \ .
\end{equation}
Our previous calculations have shown that
  the wormhole action is small when the theory allows the radial
component of the field to grow near the throat. By introducing the
exponent to the potential we were hoping to achieve several
 different purposes:

i) We were trying  to keep interactions far away from the wormhole the same as
in the usual theory of
spontaneously broken symmetry. Indeed at small   $F$ this potential coincides
with the standard potential ${1\over 4}{(F^2 - F^2_0)^2}$.

ii) When approaching small distances the value of the exponent  will change
strongly; the growth   of the field $F$ will not be able to proceed. This
should
increase the size of the   throat of the wormhole.

iii) The exponent in the potential should imitate a gravitational theory which
does not allow distances smaller than some values. In this way we could have
a model where the size of the  wormhole throat cannot be small. As we will see,
this
will give us large wormholes with the action approaching the limiting case of
the Giddings-Strominger-Lee
wormhole with the radial component of the scalar field frozen to its value at
infinity, $F = F_0$.

Our expectations were indeed confirmed by the calculations. We present the
wormhole solutions for five different values of the exponent $\beta = (0, 10,
100, 300,
500)$.
We work with $\lambda= 10^{-1}$ and $f_0 = 10^{16} $ GeV (i.e. $F_0 = 2.4
\times10^{-3}$). Fig. 5 shows
the plot of $\log_{10}F$ as the function of $\log_{10} \rho$.  At infinity all
solutions are given by the same line, corresponding to $F_0 = 2.4
\times10^{-3}$. However
they behave differently when approaching the throat.  The solutions with the
largest values  of the exponent do not grow as fast as the solution with
smaller
exponents, the grows increases when the exponent decreases. The behavior of the
radial component of the field shows the tendency to approach the GSL solution
where
the field $F$ is fixed. In our case this is a property of the solution of a
system of
equations (\ref{MAINEQ}), (\ref{MAINEQ2}), whereas the GSL solution is a
solution of
the gravitational eq. (\ref{MAINEQ2}) only, in which one assumes that $ F'=0$.
 It is
gratifying that the dynamical system with the standard mechanism of spontaneous
symmetry breaking   can be brought to the regime of the
almost frozen radial component of the filed which results in the increase of
the wormhole action
and suppression of the violation of global symmetries.
The geometric properties of the wormhole are represented in Fig. 6, which shows
the  $A(r)$ and in
Fig. 7, which shows $A'(\rho) \equiv R'(r)$. One can see a dramatic change of
the geometry with an increase of the exponent.  With zero exponent we have the
same situation as in the theory ${1\over 4}{(F^2 - F^2_0)^2}$. The wormhole
throat is very small, $A'$
becomes close to the flat space value $A' =1$ very fast. With the growth of
$\beta$ the
picture changes:  $A'$ remains smaller than $1$ even far away from $\rho=0$,
and
the size of the throat $A(0)$ becomes much larger.

The total action was calculated for several different values of the exponent
$\beta$. The results are plotted in Fig. 8.  We have found that in a wide range
of values of $\beta$  from 0 to  500 the action is given by a very simple
expression
\begin{equation}
S_{\rm total} \approx a+ b\beta \ .
\end{equation}
The values of $a,b$  for the action with the surface terms are
$a=5.7$,  $b=0.034$. The corresponding dots and the line in Fig. 9 are black.
For completeness we will also give the results for the action   without the
boundary term at the throat (see the discussion of this issue in Sec.
\ref{3s}). In this case $a=6.8$, $b=0.083$, see the grey line in Fig. 8.

Note that the value of the action
without the boundary term for the largest value of our exponent is twice as
large as the one with the boundary term, although with vanishing exponent the
difference is very small. The reason is that at higher value of the exponent
the value of the wormhole neck is getting much larger and therefore the term
${3  \pi
\over 4}  M_{\rm P}^2 R^2(0)$ is also getting larger.

We have also calculated the part of the action coming only from the area close
to the wormhole throat, where the geometry is much different from the geometry
of a flat space.  This part does not take into account the logarithmic
contribution which comes from the region with $R' \approx 1$. We have called
$S_{w}$ the part of the action which comes from the region from $r = 0$ to the
radius $r$ at which $R'(r)$ grows up to 0.9. On dimensional grounds one could
expect that $S_w$ should be a quadratic function of the size of the throat.
This is indeed the case.
 The results can be described   rather well by the following quadratic
expression:
\begin{equation}\label{quadr}
S_w = 0.16 + 85 A^2(0) \approx 0.16 + M_{\rm P}^2 R^2(0)  \ .
\end{equation}
(Note that $85 A^2 \approx M_{\rm P}^2 R^2$ for $\lambda = 0.1$). Five
dots correspond to the five values of the exponent $\beta$: the largest action
comes from the theory with
the largest exponent, see Fig. 9.

The total wormhole action (which is of main interest for us) also grows as
$M_{\rm P}^2 R^2(0)$, but with a greater coefficient:
\begin{equation}\label{quadrtotal}
S_{\rm total}  \approx 5+ 1.7   M^2_{\rm P} R^2 (0)  \ .
\end{equation}

Because of the computational difficulties, we have not performed the
calculation for the values of exponent greater than 500. At $\beta < 500$ the
logarithmic terms which are taken into account in $S_{\rm total}$ (but not in
$S_w$) also give a considerable contribution.
Therefore the quadratic fit for  $S_{\rm total}$ is much less accurate than the
fit for $S_w$, and the   coefficient $1.7$ in (\ref{quadrtotal})  is obtained
with a rather limited accuracy.
However, the behavior of the solutions  allows us to make a plausible
assumption. We expect that with the further growth of the exponential factor
the field $f(r)$ will become frozen near $f_0$, and the total wormhole action
will approach our expression for the action  (\ref{total}), (\ref{frozen})  for
the Giddings-Strominger-Lee wormhole,
\begin{equation}\label{GSLr}
S_{\rm total} = {\sqrt{3 \pi} \over 8 } \,  \Bigl(1- {2\over \pi}\Bigr)
{M_{\rm P} \over      f_0 }  = {3\pi^2
\over 8}  \left(1- {2\over \pi}\right)  M_{\rm P}^2  R^2 (0) = 1.34 M_{\rm P}^2
 R^2 (0) \ .
\end{equation}
This expression is consistent with our approximate quadratic fit
(\ref{quadrtotal}).
\vskip 0.3 cm

\section{\label{5s} Global symmetries and Planck scale physics}
\subsection{\label{5.1} Kaluza-Klein wormholes}
As we have seen, it is extremely difficult to increase the wormhole action by
changing the effective potential of the scalar field. However, as a result of
our
investigation we have learned that in those cases  when we were able to make
the action large, the value of the action could be estimated by a simple
expression $S_{\rm total}  \sim 1.34 R^2 (0)
M^2_{\rm P} $.  Remember also  that the action becomes almost three times
larger if
one does not include the inner boundary contribution.
 This suggests that one can obtain a very large action if there exists some
reason why the wormhole throat should be large. Indeed, as we have mentioned
in the Introduction, we may not have any problems with axions  if the effective
coupling constants of the operators violating the global symmetry are smaller
than $e^{-190}$. Eq. (\ref{quadr}) suggests that this  happens if the radius of
the wormhole throat becomes greater than  $15 M_{\rm P}^{-1}$. We were unable
to
make the throat that large by changing the effective potential of the scalar
field, but there exist other possibilities to do so.

Indeed, we have assumed that our space remains four-dimensional and that
gravitational interactions are described by the standard Einstein theory at all
length scales. Meanwhile, each of these assumptions may be wrong.

First of all, according to Kaluza-Klein theories, the number of dimensions of
space-time is much greater than 4, but space-time becomes effectively
four-dimensional at $R {\
\lower-1.2pt\vbox{\hbox{\rlap{$>$}\lower5pt\vbox{\hbox{$\sim$}}}}\ }R_c$, where
$R_c$ is the radius of
compactification. What if $R_c \gg M_{\rm P}^{-1}$? Then our equations should
be
considerably modified at $R {\
\lower-1.2pt\vbox{\hbox{\rlap{$<$}\lower5pt\vbox{\hbox{$\sim$}}}}\ }R_c$, which
may lead to the wormhole throat
of a large size $R(0) \sim R_c$.

It is not easy to test our hypothesis in any realistic theory, but we may play
with a toy model. First of all, at $R \gg R_c$ our equations should coincide
with our original equations (\ref{EQ}), (\ref{MAINEQQ}). Meanwhile we will
assume that at $R \ll R_c$ space-time becomes ten-dimensional, like in string
theory. In this case at $R \ll R_c$ the charge conservation equation instead of
$n = 2\pi^2 f^2 {\theta}'  R^3 $ gives $n = {\pi^2 f^2 {\theta}'  R^9 \over 12
R_c^6}$. (We assumed that 6 dimensions form a sphere $S_6$ of  radius $R_c$
and area $\pi^3R_c^3$, and took into account that the area of a  sphere $S_9$
of radius $R$ is $\pi^5 R^9\over 12$.) Modified equations equations
(\ref{EQ}), (\ref{MAINEQQ}) look as follows:
\begin{eqnarray}\label{EQ10}
f'' -{9R'f' \over R} - {dV(f) \over df}- {144 n^2 R_c^{12}\over  \pi^4 f^3
R^{18}}&= &0 \ , \\
\label{MAINEQQ10}
\nonumber\\
R'^2- 1+ {8\pi \over 3M_{\rm P}^2} R^2\left( V(f)+{72 n^2 R_c^{12} \over
\pi^4 f^2 R^{18}}-{f'^2
\over
2}\right) &=& 0\ .
\end{eqnarray}
Note also  that the factor 3 in front of ${R' f'  \over R }$ in the first of
these
equations was replaced by 9, which corresponds to space-time with $d = 10$.

The idea of a phenomenological description of   possible wormhole solutions
in this situation is to solve equations which at $R \gg R_c$ look like
(\ref{EQ}), (\ref{MAINEQQ}), but at $R \ll R_c$ look like (\ref{EQ10}),
(\ref{MAINEQQ10}). It can be achieved, e.g., by introducing an interpolating
factor $(1 + {{24}^{1/6}\,R_c\over R})^{-6}$, which changes the 9-dimensional
volume, and, correspondingly, the conservation law, $n = 2\pi^2 f^2{\theta}'
R^3  \Bigl(1 + {{24}^{1/6}\,R_c\over R}\Bigr)^{-6}$. This equation gives $n =
2\pi^2 f^2 {\theta}'  R^3 $ at $R \gg R_c$, and   $n = {\pi^2 f^2 {\theta}'
R^9
\over 12 R_c^6}$ at $R \ll R_c$. After this and some other obvious
modifications the interpolating equations can be represented in the following
way:
\begin{eqnarray}\label{EQinterp}
f''  -{3R' f'  \over R }\Bigl(1 + 2\, {{24}^{1/6}\, R_c\over R_c +R}\Bigr) -
{dV(f) \over df}- {n^2  \over 4 \pi^4 f ^3 R^{6}}\Bigl(1 +
{{24}^{1/6}\,R_c\over
R}\Bigr)^{12}&= &0 \ , \\
\label{MAINEQQinterp}
\nonumber\\
R'^2- 1+ {8\pi \over 3M_{\rm P}^2} R^2\left( V(f)+{n^2  \over
8 \pi^4 f^2 R^{6}}\Bigl(1 + {{24}^{1/6}\, R_c\over R}\Bigr)^{12}-{f'^2
\over 2}\right) &=& 0\ .
\end{eqnarray}

We have solved these equations numerically for various values of $R_c$, see
Fig. 10. As  expected, the radius of the wormhole throat $R(0)$ was found to be
approximately equal to the compactification radius $R_c$. Of course, our
investigation of Kaluza-Klein wormholes cannot be considered conclusive. Still
it indicates that the global
symmetry breaking problem may disappear in the theories where the   radius
of compactification $R_c$ is sufficiently large.

\subsection{\label{5.2} One-loop effects in quantum gravity}

Another possibility is related to the ${\cal R}^2$ corrections which may
appear in the effective Lagrangian or in equations of motion of the
gravitational field. The simplest example is   the conformal anomaly, which
gives the  contribution ${1\over6M_0^2}\ ^{(1)}H_{\mu\nu} + {1\over H_0^2}\
^{(3)}H_{\mu\nu}$ to the  gravitational equations. Here
\begin{equation}\begin{array}{rl}\label{HIJ}
^{(1)}H_{\mu\nu}\ = &{\displaystyle 2 \left(\nabla_\mu\nabla_\nu
- g_{\mu\nu} \nabla^2\right) {\cal R} + 2{\cal R} {\cal R}_{\mu\nu} -
{1\over2} g_{\mu\nu} {\cal R}^2\ ,}\\[3mm]
^{(3)}H_{\mu\nu}\ = &{\displaystyle {\cal R}_\mu^{\ \lambda}
{\cal R}_{\lambda\nu} - {2\over3} {\cal R} {\cal R}_{\mu\nu} -  {1\over2}
g_{\mu\nu} {\cal R}^{\rho\sigma} {\cal R}_{\rho\sigma} + {1\over4}
g_{\mu\nu} {\cal R}^2}\ .
\end{array}\end{equation}
The parameters $H_0$ and $M_0$ are  of the same order as $M_{\rm P}$, but  they
can be much smaller than $M_{\rm P}$ if there are many matter fields (of spin
0,
1/2,
and 1) contributing to the conformal anomaly.  Equation for the scalar field
(\ref{EQ}) remains unchanged with an account of conformal anomaly, but the
gravitation field equation (\ref{MAINEQQ}) acquires some new terms. To get an
idea of the possible influence of quantum corrections on the structure of
wormholes, let us assume for simplicity that $H_0 \ll M_0$, so that the second
term in (\ref{HIJ}) can be neglected.  In this case the gravitational equation
looks as follows:
\begin{equation}\label{EQconf}
R'^2- 1+ {8\pi R^2\over 3M_{\rm P}^2} \left( V(f)+{n^2 \over
8\pi^4 f^2 R^6}-{f'^2\over
2}\right) + {1\over R^2 H_0^2}(R'^2-1)^2  =  0 \ .
\end{equation}
This   equation on the throat yields
\begin{equation}\label{Conf}
R^2(0) = H_0^{-2} - {8\pi R^4(0)\over 3M_{\rm P}^2} \left( V(f)+{n^2 \over
8\pi^4 f^2 R^6(0)}\right)  \ .
\end{equation}

In the limit $H_0 \gg M_{\rm P}$ an additional term   ${1\over R^2
H_0^2}(R'^2-1)^2$ does not alter our wormhole solutions. However, for smaller
values of $H_0$   the character of our solution changes dramatically, see
Fig. 11. The
throat of the wormhole becomes considerably wider, and the interval of $r$
where $R' \ll 1$ becomes very small. Finally, this interval disappears
altogether, and for  $H_0 {\
\lower-1.2pt\vbox{\hbox{\rlap{$<$}\lower5pt\vbox{\hbox{$\sim$}}}}\ }2 M_{\rm
P}$ regular wormhole solutions with $R'(0) = 0$ cease to exist. In
other words, even small quantum gravity corrections can lead to absence of
wormhole solutions!

\subsection{\label{5.3} String-inspired models}

We have mentioned in Sect. \ref{3s} that the first wormhole solution was in
fact obtained in the  version of the axion theory  where instead of the
pseudoscalar axion field one has the field $H_{\mu\nu\lambda}$ \cite{GS}. In
the same paper \cite{GS} Giddings and Strominger have obtained a  family of
wormhole solutions in the string-inspired version of the theory of the field
$H_{\mu\nu\lambda}$ with the effective action
\begin{equation}\label{StringAxions}
S_{dual} = \int d^4 x \sqrt g \left(-   {\cal R} +  {1\over 2}
(\partial_\mu\phi)^2   +  e^{\beta\phi}\,H_{\mu\nu\lambda}^2
\right)-  \int_{\partial V} d^3 S (K - K_{0}) \ .
\end{equation}
Here $\phi$ is the dilaton field, $\beta$ is a phenomenological parameter;
$\beta = 2$ in string theory. For simplicity, we used here units in which
${M_{\rm P}^2 \over 16 \pi }= 1$. (Wormhole solutions in a more general class
of
theories have been obtained later by Coule and Maeda \cite{CM}). The solution
for
$R(r)$ in this theory does not depend on $\beta$. It coincides with the
corresponding solution in the theory (\ref{String}). However, the situation
with the dilaton field is more complicated. The solution can be written in the
following form:
\begin{equation}\label{StringWorm}
e^{-{\beta\over 2}\phi(r)} = e^{-{\beta\over 2}\phi(0)}\  \cos\left({\sqrt 3
\beta\over 2}\, \arccos {R^2(0)\over R^2(r)}\right) \ .
\end{equation}
For $\beta < {2\over \sqrt 3}$  this equation describes the wormhole solution
with $\phi(r)$  which gradually increases at large $r$ from its maximal value
$\phi(0)$. However, for $\beta > {2\over \sqrt 3}$ this regime becomes
impossible. Indeed, in this case $\phi(r)$ becomes infinitely large at the
point
where  ${\sqrt 3 \beta\over 2} \arccos {R^2(0)\over R^2(r)}$ becomes equal to
${\pi\over 2}$. At this point derivatives of the scalar field diverge, and the
action becomes infinitely large. The conclusion of ref. \cite{GS} was that in
such a situation there are no regular wormhole solutions in this theory.

Does it mean that the gravitational effects cannot make the field
$H_{\mu\nu\lambda}$ massive in the theory (\ref{StringAxions})? We have
already discussed a similar situation in the theory of a scalar field without
symmetry breaking, where the total action was infinite, and the conclusion was
that the effects responsible for the global symmetry violation are suppressed
only by the part of the action coming from a small vicinity of the wormhole
throat. Thus one could argue that the global symmetry violation could occur in
the theory (\ref{StringAxions}) as well, even despite the absence of the
wormhole solutions, if one considers a small part of the configuration
(\ref{StringWorm}) near the wormhole throat.

However, the main reason why this argument could work  for the theory of   a
scalar field was the existence of two vastly different length scales. The
wormhole throat had a nearly Planckian size $\sim M_{\rm P}^{-1}$, whereas the
typical scale on which the scalar field significantly changed was much greater,
of the order of $m^{-1}$. Therefore it was possible to pack many wormhole
throats inside the region of size $m^{-1}$. This is not the case for the
configurations  (\ref{StringWorm}). In the most interesting case of the string
theory with $\beta = 2$ the field $\phi(r)$ becomes infinitely large at the
point where
\begin{equation}\label{diverg}
R(r) =  {R(0)\over \cos\sqrt{\pi\over 2\sqrt 3}} \approx 1.27\, R(0) \ .
\end{equation}
Thus the total size of our field configuration almost coincides with the size
of
its throat. Since each of such configurations has infinite action and is not of
a
wormhole type, we do not expect them to lead to global symmetry breaking.

This does not necessarily imply that there are no wormhole solutions in string
theory. The model  considered above does not contain any potential
$V(\phi)$. Also, it is very hard to associate the value of the dilaton field
$\phi$, which is typically assumed to be of the order of $M_{\rm P}$,  with the
parameter $f_0 \sim 10^{12}$ GeV. Nevertheless, this model clearly shows
that the existence of wormhole solutions in the axion theory is by no means
automatic. In this model the wormhole solutions disappear as soon as one
considers effects associated with the dilaton field.

\

\section{\label{6s}Topological suppression of wormhole effects in gravity and
string theory}

All our previous results have been obtained by an investigation of
particular solutions which may or may not appear in different theories.
However, there is one general reason which may lead to suppression of
wormhole effects. These effects lead to the change of topology of space by
creating a universe capable of carrying a global charge away from our space.
On the classical level such processes simply cannot occur. Our use of
Euclidean methods to describe such processes still needs to be fully justified.
But even if these methods are valid, there is an easy  way to suppress the
probability of the wormhole formation in the Einstein theory.

In order to explain it, we remind that the standard Lagrangian of the
Yang-Mills field in QCD originally was written as ${1\over 4g^2} F_{\mu\nu}
F^{\mu\nu}$, but later it was recognized that one can add to this Lagrangian
the  term ${\theta\over 32\pi^2} F_{\mu\nu}\tilde F^{\mu\nu}$. This term
does not modify the Yang-Mills equations, but it gives a contribution to the
nonperturbative processes involving change of topology of the Yang-Mills field.

A similar situation occurs in the Einstein theory of gravity, where the
standard Einstein Lagrangian $ - {M_{\rm P}^2 \over 16 \pi } {\cal R}$ can be
supplemented by two different topological terms,
\begin{equation}
S = - \int d^4 x \sqrt g \left( {M_{\rm P}^2 \over 16 \pi } {\cal R} +
{\theta_g\over
32 \pi^2}  {\cal R}_{\mu\nu\lambda\delta} {}^*{\cal R}^{\mu\nu\lambda\delta}
+ {\gamma\over 32 \pi^2} {}^*{\cal R}_{\mu\nu\lambda\delta} {}^*{\cal
R}^{\mu\nu\lambda\delta}\right) \ .
\end{equation}
Here ${}^*{\cal R}_{\mu\nu\lambda\delta}\equiv {1\over 2} \epsilon
_{\mu\nu\mu'\nu'} {\cal R}^{\mu'\nu'}{}_{\lambda\delta}$.

The last two terms do not give any contribution to equations of motion, and
therefore do not change the theory at the classical level. Therefore, just as
the
term ${\theta\over 32\pi^2} F_{\mu\nu}\tilde F^{\mu\nu}$, these two terms
can be considered as an integral part of general relativity.   The first of
these
terms is very similar in its nature to the term ${\theta\over 32\pi^2}
F_{\mu\nu}\tilde F^{\mu\nu}$. It  contributes to the effective potential
determining the value of the axion field. Fortunately, it is not expected to
lead
to any problems with strong CP violation. The effects induced by this term
are related not to the wormhole physics, but to the Abelian Eguchi-Hanson
instantons, and typically they are exponentially small, being suppressed by
$e^{-{4\pi^2\over e^2}}$ where $e$ is the electromagnetic coupling constant
\cite{Rey}.

Meanwhile, the Gauss-Bonnet term $-{\gamma\over  32 \pi^2} {}^*{\cal
R}_{\mu\nu\lambda\delta} {}^*{\cal R}^{\mu\nu\lambda\delta}$  gives a
nonvanishing topological contribution to the wormhole action,
\begin{equation}
S_{\rm topol} = -{\gamma\over 32 \pi^2} \int d^4 x \sqrt g \ {}^*{\cal
R}_{\mu\nu\lambda\delta} {}^*{\cal R}^{\mu\nu\lambda\delta}   =-{\gamma\over 32
\pi^2}\int d^4x \sqrt g\,
({\cal R}_{\mu\nu\lambda \delta} {\cal R}^{\mu\nu\lambda \delta}
-4 {\cal R}_{\mu\nu} {\cal R}^{\mu\nu} + {\cal R}^2)\ .
\label{top}\end{equation}
Therefore it may control the strength of  the global symmetry breaking.

This term was  considered in the early works on wormholes \cite{GS}.
The constant $\gamma$ was called there a topological coupling constant. Just
as the $\theta$ parameter, the value of this constant in gravitational theory
is
not determined {\it a priory}.

It is useful to remind  the
reason why the effects of the Yang-Mills instantons are suppressed by the
factor $e^{ -{8\pi^2
\over  g^2}}$. The semiclassical Yang-Mills action ${1\over
4g^2}F_{ab} F^{ab}$ is not topological. Therefore the variation of this action
produces   equations of motion which   have
instanton solutions. The semiclassical action calculated on these solutions is
known to suppress the instantons $F_{\mu\nu} = {1\over 2} \epsilon
_{\mu\nu\lambda\delta} F_{\lambda\delta}$ as follows:
\begin{equation}
e^{-S} = e^ { - {1\over 4g^2} \int d^4 x F_{ab} F^{ab}} =
e^{ {- {8\pi^2  \over  g^2}}\nu} \ ,
\end{equation}
where $\nu$ is the winding number of the gauge configuration.

In gravity things seem to work differently,   but the results are very similar.
It is the Einstein action with matter which  gives us the equations of motions.
Those equations have wormhole solutions.
For the Euclidean wormhole geometry
(\ref{wh}) with the wormhole radius $R(r)$ the topological (Gauss-Bonnet)
contribution to the action   is given by the following integral:
\begin{equation}
S_{\rm topol} =  -{\gamma\over 32 \pi^2}\int d^4x \sqrt g\,
({\cal R}_{\mu\nu\lambda \delta} {\cal R}^{\mu\nu\lambda \delta}
-4 {\cal R}_{\mu\nu} {\cal R}^{\mu\nu} + {\cal R}^2)  =
 {3\gamma\over 4 \pi^2} \int d^4x
 { R''(r)  (1- R'^2(r))\over R^3(r)} \ .
\end{equation}

 After the angular integration the integral becomes
\begin{equation}
S_{\rm topol} =
 {3\gamma\over 2 }
\int_0^\infty  dr
  R''(r)  (1- R'^2(r)) \ .
\end{equation}
Keeping in mind that the Gauss-Bonnet part can be brought to a form where it
is a total derivative, we can rewrite this integral as
\begin{equation}\label{Top}
S_{\rm topol}  =   {3\gamma\over 2 }
\int_0^\infty  dr  {\partial\over \partial r} \left(R'(r) -  {1\over 3}
R'^3(r)\right) = {3\gamma\over 2 }
 [ R'(r) - {1\over 3}R'^3(r)]_{r\rightarrow \infty}= \gamma \ .
\end{equation}

 In performing this calculation we did not use any particular form of the
wormhole solution; we have used only the fact  that  $R' (0)=0$ and $ R'
(\infty) = 1$. Thus, independently of any suppression calculated in the
previous sections of this paper, there exists an additional exponential
suppression of the wormhole-related effects in quantum gravity by the factor
$e^{-\gamma}$. This agrees with the result obtained in \cite{GS}.

It is important that this additional suppression is equally related to the
total
probability of the wormhole formation and to the values of the vertex
operators. Indeed, for all wormhole solutions which are known to us the
derivative $R'(r)$ approaches its asymptotic regime $R'(r)= 1$ at $r$
comparable with the radius of the wormhole throat. Thus, the integral in eq.
(\ref{Top}) rapidly converges to $\gamma$ in the vicinity of the wormhole
throat. It means that all vertex operators become suppressed by the factor
$e^{-\gamma}$.

Note that the value of parameter $\gamma$ is arbitrary; it does not change any
experimentally tested predictions of general relativity. If one takes $\gamma >
190$, all our problems with axions in quantum gravity immediately disappear.
One may or may not like having a large parameter $\gamma$ in gravitational
theory, but it is certainly not forbidden, and it solves the problem of the
global symmetry breaking.

Still it would be nice to find  some reasons why this parameter should be
large. One of the ideas is to consider string theory and to study an analogous
topological term there.

In string theory it is considered plausible that the  Gauss-Bonnet term
appears   at the level of $\alpha'$ corrections in a specific form since it is
related to the Green-Schwarz mechanism of cancellation of anomalies. One
expects that this part of stringy corrections has the following form \cite{GSW}
\begin{equation}
{\cal L}_{\rm stringy} =  {\alpha' \over 16 \kappa^2} \Bigl[F_{ab} F^{ab} -
({\cal
R}_{\mu\nu\lambda \delta} {\cal R}^{\mu\nu\lambda
\delta} -4 {\cal R}_{\mu\nu} {\cal R}^{\mu\nu} + {\cal R}^2)\Bigr] \ .
\end{equation}
Here $\alpha'$ is a function of the fundamental dilaton field $\varphi$
\begin{equation}
\alpha'= {4\kappa^2 \over g_0^2} e^{ - \kappa \varphi} \ ,  \qquad \kappa^2 =
{8\pi \over M_{\rm P}^2}\ .
\end{equation}
Our normalization corresponds to the standard normalization of the
Yang-Mills Lagrangian,
\begin{equation}
{\cal L}_{\rm YM} =  {\alpha' \over 16 \kappa^2} F_{ab} F^{ab}=   {1\over
4g^2}F_{ab} F^{ab} \ ,
\end{equation}
 and we consider ${e^{ -\kappa \varphi (x)} \over g_0^2}$ as the  ``running''
gauge coupling constant ${ 1\over g^2(x)}$.

What follows is a direct generalization of our purely gravitational
calculation.
The only difference is the presence of the function $\alpha'(x) $.
Unfortunately, we do not know much about the dependence of $\alpha'$  on
$x$, which makes the results which we are going to obtain less rigorous but
perhaps still rather plausible. First of all, since we study spherically
symmetric configurations, we suppose that  $\alpha'$ depends only on  $r$ and
does not depend on the angular
variables. The integral becomes
\begin{equation}\label{strtop}
S_{\rm topol} =12\pi^2
\int_0^\infty  dr   {1\over   g^2(r)}
  R''(r)  (1-  R'^2(r))  \ .
\end{equation}
As we have seen in the previous sections, on a sufficiently large distance
$r_w$ from the wormhole throat  the wormhole geometry becomes undistinguishable
from the geometry of a flat Euclidean space because its scale factor $R(r)$
becomes almost exactly equal to $r$, and its derivative $R'(r)$ rapidly
approaches $1$. Typically it happens at the distance of the same order of
magnitude as $R(0)$; precise value of $r_w$ will not be particularly important
for us. It is important, however, that at a sufficiently large $r > r_w$ the
term $ (1-  R'^2(r))$ in the integral (\ref{strtop}) becomes very small, which
implies that the total value of the integral is dominated by integration in a
region $r < r_w$. Note also, that on all our solutions we had $(1- R'^2(r)) >0$
and $R'' > 0$. Therefore one can represent the integral (\ref{strtop})
in the following way:
\begin{equation}\label{strtop2}
S_{\rm topol}  \approx 12\pi^2
\int_0^{r_w}  dr   {1\over   g^2(r)}
  R''(r)  (1- (R'^2(r))) \equiv {12\pi^2\over g_w^2}
\int_0^{r_w}  dr
  R''(r)  (1- R'^2(r))  \ ,
\end{equation}
where $g_w$ is some average value of the gauge coupling constant in the region
$0 < r < r_w$ defined by eq. (\ref{strtop2}). Since we expect $r_w$ to be of
the same order of magnitude as $R(0)$, and   $R(0)$ should  be determined by
typical stringy length scale (just as the natural scale for $R(0)$ in the
Einstein theory was given by $M_{\rm P}^{-1}$), we will identify $g_w$ with the
typical value
of the gauge coupling constant $g_{\rm str}$ on the stringy scale. The
subsequent evaluation of the integral in (\ref{strtop2}) goes exactly as in
(\ref{Top}), because $R' \approx 1$ on the boundary $r = r_w$.
This gives
\begin{equation}\label{}
S_{\rm topol} = {8 \pi^2 \over
g_{\rm str}^2} \ .
\end{equation}
Note that this is a topological contribution to the action, which  practically
does not depend on the detailed form of the wormhole solutions (this dependence
is concentrated in our definition of $g_w \approx g_{\rm str}$).

This part of the action is a precise analog  of the
one-instanton action ${8 \pi^2 \over
g^2}$ in the Yang-Mills case. This leads to an additional suppression of the
wormhole-induced effects by the factor
\begin{equation}
e^{ - S_{\rm topol}} = e^{- {8 \pi^2 \over   g_{\rm
str}^2}} \ .
\label{supress}\end{equation}

We would like to stress that whereas in the Yang-Mills theory the suppression
of the instanton effects comes from the semiclassical action, the suppression
of
the wormhole effects described above did not came from the action of Einstein
gravity with matter but from the topological term in the action which in string
theory appears at  the level of $\alpha'$ corrections. Thus, the term $S_{\rm
topol} = {8 \pi^2 \over g_{\rm str}^2}$ appears here {\it in addition} to the
usual wormhole action.

Let us estimate the numerical value of $S_{\rm topol}$ in
the realistic theories. We do not really know the value of the effective
coupling constant  ${g_{\rm str}^2 }$ on the wormhole throat  (i.e.
approximately on the stringy scale).  The simplest idea would be to identify
${g_{\rm str}^2 }$ with  the gauge coupling constant related to the grand
unification in supersymmetric GUTs, $\alpha_{\phantom{}_{\rm GUT}} =
{g_{\phantom{}_{\rm GUT}}^2\over 4\pi} \sim {1\over 26}$ \cite{W}. This would
give $S_{\rm topol} \sim 163$.
Thus, the topological suppression alone can be strong enough to eliminate the
effects of the rank five operators $g_5 {|\Phi|^{4} (\Phi+\Phi^*)\over M_{\rm
P}}$. (Remember that the coupling constant $g_5$  for $f_0 \sim 10^{12}$ GeV
should be smaller than  $10^{-54} \sim e^{-124}$  \cite{pqgrav}.) To get the
factor
$ e^{-190}$ required to suppress the most dangerous  term
$g_1M_{\rm P}^3(\Phi + \Phi^*)$ we need a slightly smaller gauge coupling
constant,
\begin{equation}
{g_{\rm str}^2 \over 4 \pi} {\
\lower-1.2pt\vbox{\hbox{\rlap{$<$}\lower5pt\vbox{\hbox{$\sim$}}}}\ } {1\over
30}\ .
\end{equation}

Note that the factor ${8 \pi^2 \over  g_{\rm str}^2}$  is completely analogous
to
the topological coupling constant $\gamma$ which we discussed in the case of
pure gravity. In that case $\gamma$ could take any possible value, and it was
not quite clear whether it is natural or not to take it as large as $190$. In
the
case of the string theory the condition $\gamma > 190$ corresponds to a very
natural constraint ${g_{\rm str}^2 \over 4 \pi} {\
\lower-1.2pt\vbox{\hbox{\rlap{$<$}\lower5pt\vbox{\hbox{$\sim$}}}}\ } {1\over
30}$.
This requirement seems quite reasonable since the effective  gauge coupling
constant  ${g_{\phantom{}_{\rm GUT}}^2\over 4\pi} \sim {1\over 26}$ can
slightly decrease on the way from the GUT scale $2\times 10^{16}$ GeV to the
stringy scale, which is much higher. Moreover, one can solve all problems even
with   ${g_{\rm str}^2 \over 4 \pi} \sim  {1\over 26}$   if in addition to the
topological action (\ref{strtop}) one takes into account the standard
contribution to the action $S$ which we have studied in the previous sections.

It is very hard to discuss these issues in the absence of a well established
string-inspired phenomenological theory. Nevertheless we will try to make some
simple estimates. With this purpose we should remember relation between the
stringy gauge coupling constant $g_{\rm str}$, the stringy mass scale $M_{\rm
str}$ (i.e. the mass of the first massive stringy excitation), the stringy
length scale $l_{\rm str}$ (compactification radius), and the parameter
$\alpha'$ in the heterotic string theory \cite{W,Gross}:
\begin{equation}\label{PARAMETERS}
  M_{\rm str}  = {2\over\sqrt{\alpha'}} = {2\over l_{\rm str}} =
{ g_{\rm str} \, M_{\rm P}\over \sqrt {8\pi}}  \ .
\end{equation}
Here $\alpha'$ is the effective  value of  the parameter $\alpha'$ on the scale
of the wormhole throat. Therefore the topological action ${8 \pi^2 \over
g_{\rm str}^2}$ can be also written  as
\begin{equation}
S_{\rm topol} = {8 \pi^2 \over  g_{\rm str}^2} =  {\pi\over
4}
\alpha' M_{\rm P}^2 = {\pi\over 4}
 M_{\rm P}^2 l^2_{\rm str} = \pi \left(  {M_{\rm P} \over M_{\rm str}}\right)^2
\ .
\label{strtop3}
\end{equation}

Let us first concentrate on the expression $S_{\rm topol} =
\pi \left(  {M_{\rm P} \over M_{\rm str}}\right)^2$. Uncertainty in the value
of $g_{\rm str}^2$ translates into the uncertainty of  $M_{\rm str}$.  If one
simply takes $M_{\rm str} \sim  {g_{\phantom{}_{\rm GUT}} \, M_{\rm P}\over
\sqrt {8\pi}}$, one obtains $M_{\rm str} \sim1.66\times 10^{18}$ GeV, which
then leads to the estimate for the stringy unification scale  $E_{\rm str} \sim
 4 \times10^{17}$ GeV. However, this implies the existence of a large gap
between
the stringy unification scale and the SUSY GUT unification scale  $2 \times
10^{16}$ GeV. Therefore there is a tendency to assume that for some reason the
stringy scale is in fact considerably smaller than $M_{\rm str} \sim1.66\times
10^{18}$ GeV \cite{W}.  In order to obtain a sufficiently strong suppression
due to
topological effects it would be enough to have
$M_{\rm str} {\
\lower-1.2pt\vbox{\hbox{\rlap{$<$}\lower5pt\vbox{\hbox{$\sim$}}}}\ } 1.5 \times
10^{18}$ GeV, which is quite consistent with the present ideas about the value
of  $M_{\rm str}$.

Moreover, this  constraint on $M_{\rm str}$ can be somewhat  relaxed and
reduced to $M_{\rm str} {\
\lower-1.2pt\vbox{\hbox{\rlap{$<$}\lower5pt\vbox{\hbox{$\sim$}}}}\ } 2 \times
10^{18}$ GeV if one takes into account the standard (nontopological)
contribution to the wormhole action. Indeed, as we have seen, the usual
contribution to the GSL wormhole action can be represented as $1.34 M_{\rm P}^2
R(0)^2$ (\ref{GSLr}).  This result was quite consistent with our results for
the theory with the exponential potential and with simple dimensional estimates
suggesting that the action of the wormhole with the radius of the wormhole
throat $R(0)$ should be of the order $M_{\rm P}^2 R(0)^2$. Note that the
topological contribution also has the same structure (although with a slightly
smaller coefficient), $S_{\rm topol} =  {\pi\over 4}
M_{\rm P}^2 l^2_{\rm str}$. If the wormhole solutions exist at all in the
string theory, one may expect that the wormhole throat $R(0)$ should be greater
than the ``elementary length scale'' $l_{\rm str}$. This suggests that the
nontopological part of the  action $\sim M_{\rm P}^2 R(0)^2$ should be greater
than $M_{\rm P}^2 l^2_{\rm str}$. If this is correct, the total action of the
wormholes, including the topological part, should be about two times greater
than the topological contribution. In fact, if one simply adds to the
topological action $ {\pi\over 4}  M_{\rm P}^2 l^2_{\rm str}$ the GSL action
$1.34 M_{\rm P}^2 R(0)^2$ for $R(0) \sim l_{\rm str}$, the value of the
topological action almost triples. In such a  situation one can expect that the
total action should become greater than  400 even if one takes ${g_{\rm str}^2
\over 4 \pi} \sim    {g_{\phantom{}_{\rm GUT}}^2\over 4\pi  } \sim {1\over 26}
$. This is more than sufficient to solve the problem of the global symmetry
violation. One would get the total action $S {\
\lower-1.2pt\vbox{\hbox{\rlap{$>$}\lower5pt\vbox{\hbox{$\sim$}}}}\ }190$ for
${g_{\rm str}^2 \over 4 \pi} {\
\lower-1.2pt\vbox{\hbox{\rlap{$<$}\lower5pt\vbox{\hbox{$\sim$}}}}\ }  {1\over
15} $, which looks like a very safe bet.

Thus, instead of the absolutely incredible fine-tuning of the values of
possible coupling
constants of the interaction terms breaking the global symmetry in the axion
theory, our estimates gave us rather mild constraints on the gauge coupling
constant and on the stringy mass scale $M_{\rm str}$. This provides a natural
possibility to make stringy gravity compatible
with the existence of the light axion in Nature.

\vfill
\newpage

\section{\label{7s} Discussion}

One of the main obstacles on the way of development of quantum gravity is the
problem of experimental verification of its predictions. Indeed, gravitational
interactions between elementary particles become strong only at energies
comparable to the Planck mass, $M_{\rm P} \sim 1.2\times 10^{19}$ GeV.  Such
energies are far beyond our reach.

Fortunately, there exist some  indirect ways to test quantum gravity
experimentally. For example, it is quite possible that a consistent theory of
gravity requires supersymmetry. Then one may study different versions of the
theory by investigation of the properties of light superpartners of the
graviton. Another possibility is related to  cosmology, which provides us with
experimental
data originated at the very early stages of the evolution of the Universe.

In addition to that, there may exist some nonperturbative gravitational effects
which may have important experimentally testable consequences even at very
low energies. Some of these effects have been investigated in this paper. We
have found, under the assumptions specified in this paper, that gravitational
effects strongly violate global symmetries in a wide class of theories,
including the theory of axions. Surprisingly enough, this strong violation
occurs even if one drastically modifies the effective potential of the theory,
for example, if one multiplies it by $\exp{C\Phi\over M_{\rm P}}$, where the
factor
$C$ can be as big as $10^3$.

Of course, it is quite possible that our methods based upon Euclidean
approach to quantum gravity are inadequate. However, we must admit that when
we began our investigation we expected that it will be very easy to fix this
problem either by using a formulation where the global symmetries become
local, or by finding a simple modification of the theory which leads to
wormholes with a very large action.  We have found that it is almost impossible
to do so in the context of the standard Einstein theory in a four-dimensional
space.

However, as soon as we allowed ourselves to modify  the theory of gravity or
the properties of space at the scale about $10 M_{\rm P}^{-1}$ we have found
many different possibilities to improve the situation. One of them is the
possibility that our space is compactified, with a compactification radius $r_c
{\ \lower-1.2pt\vbox{\hbox{\rlap{$>$}\lower5pt\vbox{\hbox{$\sim$}}}}\ } 10
M_{\rm P}^{-1}$. We have seen also that wormholes may simply disappear if one
takes into account conformal anomaly, or if one considers certain
string-inspired axion models. In addition to all these effects, we have found
that in  string theory there exists a specific strong suppression of topology
change by the factor $e^{- {8 \pi^2 \over   g_{\rm
str}^2}} = e^{\pi \left(  {M_{\rm P} \over M_{\rm str}}\right)^2}$. This is a
topological effect which does not depend on many particular details of
wormhole configurations. Our estimates show that with an account taken of
topological terms, the effects related
to the global symmetry breaking become strongly suppressed if the string mass
scale is sufficiently small, $M_{\rm str} {\
\lower-1.2pt\vbox{\hbox{\rlap{$<$}\lower5pt\vbox{\hbox{$\sim$}}}}\ } 2 \times
10^{18}$ GeV. Equivalently,  the problem of the wormhole-induced global
symmetry breaking disappears if the gauge coupling constant is sufficiently
small on the stringy scale of energies, ${g_{\rm str}^2 \over 4 \pi} {\
\lower-1.2pt\vbox{\hbox{\rlap{$<$}\lower5pt\vbox{\hbox{$\sim$}}}}\ } {1\over
15}$.
These values of parameters are quite consistent with the existing picture of
stringy phenomenology. Therefore at present we do not see
any reason to reject the theories with global symmetries.

On the other hand, we have found that the existence of (approximate) global
symmetries and the possibility to solve the strong CP violation problem by the
Peccei-Quinn mechanism are very sensitive to the choice of the theory of
quantum gravity.   If  the axions in the mass range of  $m_a \sim 10^{-5}$ eV
will be discovered  experimentally \cite{Exper}, it may give us   important
information about the structure of space and properties of particle
interactions at the
Planckian scale.

\section*{Acknowledgments}

It is a pleasure to thank  Michael Dine  for many enlightening discussions at
different stages of this investigation. This work was supported in part by NSF
Grant  No. PHY-8612280.

\vfill
\newpage

\section*{\label{8s} Appendix:  Numerical Methods of Finding Wormhole
Solutions}
Finding   the wormhole solutions numerically was a rather complicated problem.
We have found that the behavior of the solutions was extremely sensitive to the
 choice of initial conditions. Whereas it was relatively easy to find solutions
with small action $S$, in the most interesting cases where the action was large
the standard numerical methods failed. Therefore it was necessary to develop a
more advanced method of calculations.  There is a chance that our method  can
be useful for solving other problems as well. Therefore in this Appendix we
will first describe the standard  numerical methods of obtaining   solutions to
the differential equations
(\ref{MAINEQ}), (\ref{MAINEQ2}), and then we will describe our method.

\subsection*{Standard Shooting Method}
The first method we try is the standard shooting method \cite{PTVF}. We fix
some value for
$F(0)$. Then we find $A(0)$ using (\ref{COND1}). This involves solving a cubic
equation for $A^2(0)$.
\begin{equation}
\label{INIEQ}
A^4(0)-A^6(0)\,V(F(0))-{Q \over F^2(0)}=0 \ .
\end{equation}
If there are no roots, then eqs. (\ref{MAINEQ}), (\ref{MAINEQ2}) do  not have a
solution for the fixed value of $F(0)$. Eq. (\ref{INIEQ}) has at most two
roots. We find that the bigger root does not lead to a solution satisfying
boundary conditions at infinity. Hence we pick the smallest root.

Once we have the initial conditions we use any numerical ordinary differential
equation solver (for example Runge-Kutta  method). According to the behavior of
the solution at infinity we modify our guess of $F(0)$. Using a simple
iterative procedure we can determine the correct value for $F(0)$.

 First we have used Runge-Kutta  method to obtain the numerical solution once
the initial conditions (boundary conditions at $x=0$) are set. However, for
solutions with large action, the solution has to be computed to a very high
accuracy. We have found that the Runge-Kutta  method does not work for actions
above 6.
Bulirsch-Stoer method, which we used next, works for actions below 10. As a
result, simple shooting method is inadequate for our purposes. The reason it
does not work is that $F(0)$ has to be determined to a very high accuracy,
otherwise the numerical solution will not come close to the boundary conditions
at infinity (at $x\approx {10 \over F_0}$). On a computer, $F(0)$ can be
determined only to machine precision. The difference between $F(0)$ rounded to
machine precision and the correct $F(0)$ causes an exponentially large
deviation between the numerical solution and the correct solution for large
$x$,
hence making the standard shooting method fail.

\subsection*{Improved Shooting Method}
With the standard shooting method we cannot get a numerical solution to closely
approximate the correct solution for the wormholes with large action. The idea
of
an alternative method is to sacrifice the correctness of the numerical solution
at some $x$, and use the gained freedom to make it closer to the correct
solution. The disadvantage is that the resulting numerical solution does not
satisfy the differential equation at all points. The advantage is that it is
close to the correct solution.

We solve the differential equation in several stages. The first stage uses the
standard shooting method with the Bulirsch-Stoer method as the ordinary
differential
equation solver. After $F(0)$ has been determined within machine precision, the
first stage is completed. Every time a solution is generated with the
Bulirsch-Stoer method, the program remembers at which $x$  the solution rapidly
veered up or down so that it is clear it will not satisfy the boundary
conditions at infinity. Once the stage is completed, the program recalls
solutions $F_{\rm up}(x)$ and $F_{\rm down}(x)$ which are the curves which have
passed next to the correct solution for the longest time. $F_{\rm up}(x)$ has
passed above the correct solution (so it veered up) and $F_{\rm down}(x)$
passed below the correct solution.

The  program finds the maximum $x$ for which the two solutions are still very
close to each other: $|{F_{\rm up}(x)-F_{\rm down}(x)}|<\epsilon$. Let this
point be $x_m$. Then we have found a good approximation to the correct solution
up to $x=x_m$. The second stage is to try to find an approximation to the
solution for $x>x_m$. For the second stage, and all further stages, the initial
conditions are different from the ones at the first stage. As in the case of
the first stage, we are free to choose $F(x_m)$ with the restriction $F_{\rm
up}(x_m)\geq F(x_m)\geq F_{\rm down}(x_m)$. Notice that by choosing $F(x_m)$ we
introduce a point in the numerical solution which does not satisfy the
differential equation. The error introduced is very small, though, since
$F_{\rm up}(x_m)$ is very close to $F_{\rm down}(x_m)$.

We also need to know $A(x_m)$. We find it by linear interpolation between
$A_{\rm up}(x_m)$ and $A_{\rm down}(x_m)$ once $F(x_m)$ is set. In the same way
we determine $F'(x_m)$. As already mentioned, the jump in $F(x), A(x), F'(x)$
at
$x=x_m$ is small and can be made as small as we like by reducing the parameter
$\epsilon$. During the second stage $F(x_m)$ is determined to machine precision
by the same iterative technique used in the first stage to determine $F(0)$.
Then we proceed to the third stage, and so on, until we reach $x\approx  {10
\over F_0}$ after which the solution is nearly constant: $F(x)\sim F_0, \;
x\rightarrow \infty$.

The method described above can be interpreted in a way similar to the
interpretation
of the standard shooting method. We aim and shoot for the boundary conditions
at infinity. Having missed several times, we follow the trajectory the best
arrows followed and approach the target a little closer. Then we shoot from the
new
position. After finally hitting the target we present the set of trajectories
of different arrows as one nearly correct trajectory which a single arrow could
have followed all the way from the initial position to the target. Being a  bad
shot, we would not have been able to shoot the arrow to follow the correct
trajectory all the
way from the boundary   at $x=0$ to the boundary  at infinity.
However, using the method described above we can find a good approximation to
the correct arrow
trajectory.

\vfill
\newpage

\section*{Figure Captions}

{}~~~~~ Figure 1:  (a) The geometry of the Giddings-Strominger-Lee wormhole
with
initial surface $\Sigma_i$ which is $R^3$ and final surface
$\Sigma_f$ which is $R^3\times S^3$. This wormhole may describe the tunneling
from
$R^3$ to $R^3\times S^3$.
{}~(b) Extended
solution which connects two
asymptotically Euclidean regions $r\rightarrow \pm \infty$.

 Figure 2: The distribution of $\log_{10} F$ as a function of $\log_{10} \rho$
for the wormhole solution in the theory with an effective potential
${\lambda\over 4} (f^2-f_0^2)^2$.  Here $ \rho  = r M_{\rm P} \sqrt {  3\lambda
\over 8\pi}$, $F= {f\over M_{\rm P}} \sqrt {8\pi   \over 3}$.

 Figure 3: Wormhole geometry for the potential ${\lambda\over 4} (f^2
-f_0^2)^2$. Here $\rho  = r M_{\rm P}  \sqrt {  3\lambda \over 8\pi}$, $A = R
M_{\rm P}  \sqrt {  3\lambda \over 8\pi}$.

Figure 4: Total action as a function of $- \ln F_0 \equiv \ln {M_{\rm P}\over
f} + {1\over 2} \ln {3\over 8\pi}$ for $\lambda = 0.1$. The black dots and the
line show  the total action  with the boundary term at the throat, the grey
ones give
the corresponding values for the action without the boundary term. Note that
$S_{\rm total} \sim 15$ for $f_0 \sim 10^{12}$ GeV.

 Figure 5: Plot of $\log_{10} F$ for different values of the exponent  and
$f_0
= 10^{16} $ GeV .

 Figure 6: Scale factor $A(\rho)$ for different values of the exponent and $f_0
= 10^{16}$ GeV.

 Figure 7: Derivative of the scale factor $A(\rho)$
 for different values of
the exponent and $f_0 = 10^{16}$ GeV. Wormhole geometry approaches geometry of
a flat Euclidean space when  $A' (\rho)$ approaches 1.

 Figure 8: The total action in the theory with the exponential potential $V_3
(f)=\frac{\lambda}{4}e^{\beta fM^{-1}_{\rm P}}\left( f^2-f_0^2 \right)^2$ as a
function of the exponent $\beta$. Black dots give the action with the boundary
term, grey one give the action without the boundary term at the wormhole
throat.

 Figure 9: The action obtained by integration in the region near the wormhole
throat  (where $A'(\rho)=R'(r) < 0.9$) for different values of the exponent
$\beta$.

 Figure 10: Behavior of the radius $R$ of the wormhole solution at small $r$ in
our model of a $10$-dimensional Kaluza-Klein theory with different values of
the compactification radius $R_c$. All quantities are given in the Planck units
$M_{\rm P}^{-1} = 1$.

 Figure 11: Behavior of the radius $A(\rho)$ of the wormhole solution at small
$\rho$ in the model taking into account conformal anomaly. Note that  regular
wormhole solutions with $A'(0) = 0$  (i.e. with $R'(0) = 0$) disappear for
$H_0 {\
\lower-1.2pt\vbox{\hbox{\rlap{$<$}\lower5pt\vbox{\hbox{$\sim$}}}}\ }2 M_{\rm
P}$, even though in this regime the coefficient ${1\over H_0^2}$ in front of
the conformal anomaly term is still very small.

\end{document}